\documentclass[12pt]{article}

\usepackage{latexsym}
\usepackage{epsf}
\usepackage{epsfig}
\usepackage{amssymb,amsmath}
\usepackage[english]{babel}
\usepackage{amsfonts}
\usepackage{amssymb}
\usepackage{graphics}
\usepackage{mathrsfs}
\usepackage{verbatim}
\usepackage{float}
\def\beq{\begin{equation}}
\def\eeq{\end{equation}}
\def\bea{\begin{eqnarray}}
\def\eea{\end{eqnarray}}
\newcommand{\bal}{\begin{align}}
\newcommand{\eal}{\end{align}}
\def\ba{\begin{array}}
\def\ea{\end{array}}
\def\bi{\begin{itemize}}
\def\ei{\end{itemize}}
\def\ben{\begin{enumerate}}
\def\een{\end{enumerate}}
\def\beq{\begin{equation}}
\def\eeq{\end{equation}}
\def\bc{\begin{center}}
\def\ec{\end{center}}
 \def\bt{\begin{table}}
\def\et{\end{table}}
 \def\btb{\begin{tabular}}
\def\etb{\end{tabular}}

\def\mass2{mass${}^2$}

\def\simlt{\stackrel{<}{{}_\sim}}
\def\simgt{\stackrel{>}{{}_\sim}}

\title{
\begin{flushright}
\normalsize{ ANL-HEP-PR-09-15\\ EFI-09-08\\}
\end{flushright}
\vspace*{5mm} \Large\textbf{A Heavy Higgs and a Light Sneutrino NLSP in the MSSM with Enhanced SU(2) D-terms} \vspace*{0.5cm}
\author{\textbf{Anibal D.~Medina~$^{a}$, Nausheen
R.~Shah~$^{b,d}$}\\ \textbf{and Carlos E.~M.~Wagner~$^{b,c,d}$}\\
\\
\normalsize\emph{Department of Physics, University of California, One Shields Ave. Davis, CA 95616~$^a$, USA}\\ \normalsize\emph{Enrico Fermi
Institute~$^b$ and Kavli Institute for Cosmological Physics~$^c$,}\\ \normalsize\emph{University of Chicago, , 5640 S. Ellis Ave., Chicago, IL 60637,
USA} \\ \normalsize\emph{HEP Division, Argonne National Laboratory, 9700 Cass Ave., Argonne, IL 60439, USA~$^d$}}}

\date{\today}
\begin{document}
\setcounter{page}{0} \maketitle

\vspace*{0.5cm} \maketitle 
\begin{abstract}
The Minimal Supersymmetric extension of the Standard Model provides a solution to the hierarchy problem and leads to the presence of a light Higgs. A
Higgs boson with mass above the present experimental bound may only be obtained for relatively heavy third generation squarks, requiring a precise,
somewhat unnatural balance between different contributions to the effective Higgs mass parameter. It was recently noticed that somewhat heavier Higgs
bosons, which are naturally beyond the LEP bound, may be obtained by enhanced weak $SU(2)$ D-terms. Such contributions appear in models with an
enhanced electroweak gauge symmetry, provided the supersymmetry breaking masses associated with the scalars responsible for the breakdown of the
enhanced gauge symmetry group to the Standard Model one, are larger than the enhanced symmetry breaking scale. In this article we emphasize that the
enhanced $SU(2)$ D-terms will not only raise the Higgs boson mass but also affect the spectrum of the non-standard Higgs bosons, sleptons and squarks,
which therefore provide a natural contribution to the $T$ parameter, compensating for the negative one coming from the heavy Higgs boson. The sleptons
and non-standard Higgs bosons of these models, in particular, may act in a way similar to the so-called inert Higgs doublet. The phenomenological
properties of these models are emphasized and possible cosmological implications as well as collider signatures are described.
\end{abstract}

\thispagestyle{empty}

\newpage

\setcounter{page}{1}

\section{Introduction}

The Minimal Supersymmetric Standard Model (MSSM) provides an
elegant solution to the hierarchy problem of the Standard Model
(SM) and in the presence of R-parity, naturally includes a good
dark matter candidate. The MSSM also leads to the presence of a
Higgs boson, with SM-like properties, which is naturally light,
with a tree-level mass smaller than the $Z$-gauge boson
mass~\cite{Haber:1984rc,Martin:1997ns}. Therefore, the current
experimental bound on a SM-like Higgs boson from the LEP
experiment, $m_h \simgt 114.4$~GeV, demands relatively large
radiative corrections induced by the third generation
squarks~\cite{Haber:1990aw},\cite{Carena:1995bx}. Such large
corrections may only be obtained by large squark masses which lead
at the same time to large negative radiative corrections to the
soft supersymmetry breaking Higgs mass parameter. Hence, proper
electroweak symmetry breaking (EWSB) may only be obtained by
fine-tuning the negative soft supersymmetry breaking and positive
$\mu$ term contributions to the effective Higgs mass parameter.

The situation may be improved if the electroweak SM gauge group
proceeds from a larger gauge structure which is spontaneously
broken to the SM group at energies of a few TeV. If the
supersymmetry breaking parameter of the scalars, which are
associated with the breakdown of the extended gauge symmetry, is
much larger than the order parameter of the gauge symmetry
breakdown, the D-term of the low energy weak interactions may be
enhanced. This, in turn, leads to larger values of the tree-level
SM-like Higgs boson mass and reduces fine
tuning~\cite{Batra:2003nj},\cite{Batra:2004vc},\cite{Bellazzini:2009ix}.

A natural implementation of these ideas may be realized by
extending the weak gauge group to the group $SU(2)_1 \times
SU(2)_2$, where the two Higgs doublets of the MSSM as well as the
third generation particles are charged under one of the two
$SU(2)$ groups, $SU(2)_1$, while the first and second generation
particles are charged under $SU(2)_2$~\cite{Muller:1996dj}. The
$SU(2)_1 \times SU(2)_2$ symmetry is eventually broken to the
diagonal $SU(2)_W$ group that may be identified with the one
associated with the SM weak interactions. Such an extension of the
MSSM is anomaly free and can lead to an understanding of the
hierarchy of the third generation and the lighter generation
masses. For relatively large values of the $SU(2)_1$ gauge
coupling one can get SM-like Higgs boson masses up to 300~GeV.

Such large Higgs boson masses tend to be in conflict with
precision electroweak observables~\cite{Amsler:2008zzb}. The model
described above, provides extra contributions to these
observables. On one side, there are non-universal contributions to
the third generation quark and lepton couplings to the weak gauge
bosons, which are induced by the mixing of the heavy gauge bosons
with the SM ones. Such contributions may be minimized for large
values of the vacuum expectation value (vev) of the scalars that
provide the breakdown of the extended symmetry (see, for example,
Ref.~\cite{Morrissey:2005uza}). However, the presence of a heavy
Higgs still leads to a relatively large negative contribution to
the $T$-parameter, leading to further tension with the precision
electroweak observables.

In this work we emphasize that the same D-term contributions that
lead to an enhancement of the Higgs mass also lead to a splitting
of the left-handed super-partners of the third generation squarks
and sleptons as well as the non-standard Higgs bosons. We show
that the mass splitting of these doublets is naturally of the size
necessary to compensate the large negative Higgs contribution,
restoring agreement of the model with precision electroweak data,
similar to the inert Higgs doublet models~\cite{Barbieri:2006dq}.
The modification of the spectrum leads to novel phenomenological
properties with respect to the MSSM. It is the aim of this article
to discuss these properties as well as the associated
supersymmetry breaking mechanism that may be consistent with such
a spectrum.

The article is organized as follows. In section~2 we review the
model. In section~3 we describe the Higgs and sparticle spectrum
and discuss the contributions to the $T$-parameter. In section~4
we describe a possible realization of supersymmetry breaking that
leads to a spectrum of third generation left-handed sparticles
much lighter than the first and second generation ones and discuss
the collider physics associated with this model. In section~5 we
comment on the possible cosmological constraints. In section~6 we
describe a different possibility, in which the non-standard Higgs
bosons become light, while the left-handed squarks and sleptons
remain heavy. We reserve section~7 for our conclusions.

\section{Review of the Model}

A simple way to implement the possibility of an enhanced
electroweak D-term is through the enlarged weak gauge group
$SU(2)_1 \times SU(2)_2$. The two MSSM Higgs bosons and the third
generation left-handed doublet super-fields are charged under
$SU(2)_1$ but singlets under $SU(2)_2$, while the lighter
generations left-handed doublets are singlets under $SU(2)_1$ but
charged under $SU(2)_2$. A bidoublet, $\Sigma$, is introduced
which acquires a vev, $<\Sigma> = u I$, breaking the product group
to the diagonal one, $SU(2)_W$, which is identified with the SM
weak group. The effective weak gauge coupling is therefore given
by
\begin{equation}
g = \frac{g_1 g_2}{\sqrt{g_1^2 + g_2^2}},
\end{equation}
where $g_{1}$ and $g_2$ are the $SU(2)_1$ and $SU(2)_2$ gauge
couplings, respectively. For values of $g_1 \gg g_2$ the weak
coupling is approximately equal to $g_2$. Following
Ref~\cite{Batra:2003nj}, the breakdown of the enhanced gauge
symmetry is governed by the $\Sigma$ superpotential
\begin{equation}
W = \lambda_1 S \left(\frac{\Sigma \Sigma}{2} - w^2 \right),
\end{equation}
where $S$ is a gauge singlet. This leads to a $\Sigma$ potential
of the form
\begin{equation}
V =  m_{\Sigma}^2 \Sigma^{\dagger} \Sigma  + \frac{\lambda_1^2}{4} \left| \Sigma \Sigma \right|^2 - \frac{B}{2} \left(\Sigma \Sigma + h.c.\right) + ...
\label{potential}
\end{equation}
where $B = \lambda_1 w^2$, $m_{\Sigma}^2$ is a soft supersymmetry breaking mass term and, for simplicity, we have considered all parameters to be real. There is also a contribution coming from the D-terms of the $SU(2)_1$ and $SU(2)_2$ gauge groups to the potential,
\begin{equation}
\Delta V = \frac{g_{1}^2}{8}\left( \rm{Tr}[ \Sigma^{\dagger} \tau^a \Sigma ] + H_u^{\dagger} \tau^a H_u + H_d^{\dagger} \tau^a H_d + L^{\dagger} \tau^a
L + Q^{\dagger} \tau^a Q \right)^2 + \frac{g_{2}^2}{8}\left( \rm{Tr}[ \Sigma^{\dagger} \tau^a \Sigma ] + ... \right)^2
\end{equation}
where $\tau^{a}$ are the generators of $SU(2)$ and, for
completeness, we have added the extra third generation lepton and
quark scalars, which were omitted in the analysis of
Ref.~\cite{Batra:2003nj}.

For values of $B > m_{\Sigma}^2$, the $\Sigma$ field acquires a
vev along the D-flat direction, $<\Sigma> = u I$, with $u^2 = (B -
m^2_{\Sigma})/ \lambda_1^2$. Assuming that $B \gg v^2$, with $v$
the vev of the SM Higgs boson, one can integrate out the heavy
degrees of freedom and find the result for the enhanced $SU(2)_W$
D-terms
\begin{equation}
\Delta V = \frac{g^2}{2} \Delta \sum_a \left( H_u^{\dagger} \tau^a H_u + H_d^{\dagger} \tau^a H_d + L_3^{\dagger} \tau^a L_3 + Q_3^{\dagger} \tau^a Q_3
\right)^2
\end{equation}
and
\begin{equation}
\Delta = \frac{1 + \frac{2 m^2_{\Sigma}}{g_2^2 u^2}}{1 + \frac{2 m^2_{\Sigma}} {(g_2^2 + g_1^2) u^2}}.
\end{equation}
Therefore, ignoring mixing with the non-standard CP-even Higgs
boson, the SM-like Higgs mass is enhanced to a value
\begin{equation}
m_h^2 = \frac{1}{2} \left(g^2 \Delta + g_Y^2 \right) v^2 \cos^2 2 \beta + \lambda_2^2 v^2 \sin^2 2 \beta + {\rm loop \; corrections}
\end{equation}
which is the main result of Ref.~\cite{Batra:2003nj}. For
completeness, we added the possible contribution of a
superpotential term
\begin{equation}
W_{\lambda} = \lambda_2 S' H_u H_d
\end{equation}
with $S'$ a singlet superfield. In our analysis we shall assume
that the D-terms give the dominant contribution to the tree-level
Higgs mass. The term $\lambda_2$, however, may have important
phenomenological implications as have been observed in several
works, including Refs.~\cite{Batra:2004vc,Barbieri:2006bg}. In our
phenomenological analysis, we will consider the simple case in
which $\lambda_2 = 0$.

Assuming that the value of $u$ is of the order of a few TeV, as
required to minimize the non-oblique corrections to precision
electroweak observables, the authors of Ref.~\cite{Batra:2003nj}
showed that the presence of the enhanced D-terms can naturally
raise the Higgs mass up to values of about 250~GeV, and for large
values of the $g_1$ coupling at the symmetry breaking scale,
$\alpha_1 \simeq 1$, values as large as 300~GeV may be obtained.

\section{Sparticle Spectrum and Phenomenological Implications}

An important aspect of this model that was omitted from
Ref.~\cite{Batra:2003nj} is that the presence of the D-terms will
also induce additional splitting in the Higgs, squark and slepton
masses. In order to understand this point, one can write the low
energy $SU(2)$ D-term contributions to the effective potential in
the following way
\begin{eqnarray}
V_D = \frac{g^2 \Delta}{8} \left( \sum_i \Phi_i^{\dagger}\Phi_i \right)^2 - \frac{g^2 \Delta}{4} \sum_{ij} \left| \Phi_i^{T} i \sigma_2 \Phi_j
\right|^2, \label{VD}
\end{eqnarray}
where $\Phi_i = L_3,~Q_3,~H_i$ are the $SU(2)$ doublets with
non-trivial transformations under $SU(2)_1$ and $\sigma_2$ is the
antisymmetric Pauli matrix. To obtain the spectrum one should
concentrate on interactions of the different particles with the
Higgs fields. The first term in Eq.~(\ref{VD}) leads to a positive
contribution, $g^2 \Delta v^2/4$ to the mass of the third
generation squarks and sleptons, where $v^2 = v_u^2 + v_d^2$,~and
$v_{u,d}$ are the $H_{u,d}$ vevs. The second term in $V_D$ is the
sum of the squares of the $SU(2)$ invariant combinations that
couple to the corresponding quark and lepton $SU(2)$ singlets in
the superpotential Yukawa terms. Therefore, it leads to negative
contributions, $-g^2 \Delta v_i^2/2$, to the squark and slepton
doublet component that couples via the Yukawa interaction with the
Higgs $H_i$, with $i=u,d$.

The sparticles also receives $F$-term contributions associated
with the superpotential
\begin{equation}
W = \mu H_u H_d  + h_u H_u Q U + h_d H_d Q D + h_l H_d L E .
\end{equation}
In particular, the third generation squarks receive supersymmetric
contributions proportional to their super-partner masses, which
become particularly important in the case of the top squarks. The
combination of the F-term and D-term contributions lead to the
following splitting between the third generation left-handed
squark and slepton mass parameters,
\begin{eqnarray}
m_{\tilde{\tau}_L}^2 - m_{\tilde \nu_{\tau}}^2 & = & \Delta_D \\ m_{\tilde{b}_L}^2 - m_{\tilde{t}_L}^2 & = & \Delta_D - m_t^2\label{topbottom}
\end{eqnarray}
and
\begin{eqnarray}
\Delta_D & = & \frac{g^2 v^2}{2} \Delta |\cos 2 \beta| \nonumber\\ & = & \left(\Delta m_h^2 \right)_D /|\cos 2 \beta|
\end{eqnarray}
where $v \simeq 174.1$~GeV is the SM Higgs boson vev and $(\Delta
m_h^2)_D = g^2 v^2 \Delta \cos^2 (2 \beta)/2$ is the enhanced
D-term contribution to the tree-level Higgs mass and we have
ignored all SM quark and lepton masses apart from the top quark
mass. For moderate to large values of $\tan\beta$, $\tan\beta >
5$, which we will assume from now on, $\Delta_D \simeq (\Delta
m_h^2)_D$.

The case of the non-standard Higgs bosons should be treated
separately. The charged Higgs and CP-odd states are defined as
orthogonal to the corresponding charged and neutral Goldstone
boson states, namely
\begin{eqnarray}
H^{\pm} & = & \sin\beta \; H_d^{\pm} - \cos\beta  \; H_u^{\pm} \nonumber\\ A   & = & \sin\beta \; A_d - \cos\beta \; A_u
\end{eqnarray}
where $H_i^\pm$ and $A_i$ are the charged Higgs and CP-odd
components of the Higgs field doublets $H_i$ ($i=u,d$), and for
$H_u$ we redefine $H_u \rightarrow i \sigma_2 H_u^{*}$ which
carries the same hypercharge as $H_d$.

Similar to the MSSM case, the neutral CP-odd mass is given by
\begin{equation}
m_A^2 = m_1^2 + m_2^2 
\end{equation}
where $m_1^2 = m_{H_d}^2 + |\mu|^2$ and $m_2^2 = m_{H_u}^2 +
|\mu|^2$, where $m_{H_d}^2$ and $m_{H_u}^2$ are the soft
supersymmetry breaking square mass parameters. The resultant mass
splitting is given by
\begin{equation}
m_{H^\pm}^2 - m_A^2 =  \frac{g^2 \Delta}{2} v^2.
\end{equation}

\subsection{Contributions to the $T$-Parameter}

The additional splitting described in the previous section between
the masses of the Higgs bosons, sleptons and squarks can have
important phenomenological implications. For instance, contrary to
naive expectations, for large values of the D-term contribution to
the Higgs mass, the left-handed sbottom becomes heavier than the
left-handed stops. The exact stop and sbottom spectrum depends,
however, on the mixing with the right-handed third generation
squarks. In this section, to simplify the discussion, we will
assume those mixings to be small, something that is natural for
non-degenerate squarks, moderate values of $A_t$ and non-extreme
values of $\tan\beta$.

The splitting between the upper and lower components of the
left-handed doublets control their contribution to the $T$
parameter. For an $SU(2)$ doublet, with up and down mass
eigenvalues $m_u$ and $m_d$, the contribution to the $T$-parameter
is given by
\begin{eqnarray}
\Delta T & = & \frac{N_c}{16 \pi s_W^2 m_W^2} \left[ m_u^2 + m_d^2 - \frac{2 m_u^2 m_d^2}{m_u^2 - m_d^2} \log\left( \frac{m_u^2}{m_d^2} \right)
\right],
\end{eqnarray}
where $s_W$ is the sine of the standard weak mixing angle $s_W^2
\simeq 0.2315$.

If the ratio of the heavier doublet mass eigenstate to the lighter
one is smaller than about 3, the contribution to the $T$ parameter
may be approximated by~\cite{Barbieri:2006dq}
\begin{eqnarray}
\Delta T & = & \frac{N_c}{12 \pi s_W^2 m_W^2} (\Delta m_{ud})^2 \nonumber\\ & = & \frac{N_c}{12 \pi s_W^2 m_W^2} \frac{(\Delta m^2_{ud})^2}{(m_u +
m_d)^2},\label{deltaTsimple}
\end{eqnarray}
where $N_c$ is the number of colors, $\Delta m_{ud}$ and $\Delta m^2_{ud}$ are the differences between the masses and the squared masses of the heavier and lighter components of the doublet. These contributions should be added to the ones associated with the heavy SM-like Higgs boson,
\begin{eqnarray}
\Delta T & = & - \frac{3}{8 \pi c_W^2} \ln \frac{m_h}{m_{h_{\rm ref}}} \nonumber\\ \Delta S & = & \frac{1}{6 \pi} \ln \frac{m_h}{m_{h_{\rm ref}}},
\end{eqnarray}
where $m_{h_{\rm ref}}$ is a Higgs mass reference value. As
mentioned above, the up and down sfermion mass eigenstates are
admixtures of $SU(2)$ doublet and singlet components. In the case
of non-negligible mixing, the expression for $\Delta T$, given by
Eq.~(\ref{deltaTsimple}), needs to be reformulated. This turns out
to be specially important for third generation sfermions. In the
case of an $SU(2)$ doublet, taking into account mixing angles, the
general expression for $\Delta T$ takes the form,
\begin{eqnarray}
\Delta T&=&\frac{N_c}{12 \pi s^2_W m_W^2} \left(\sin^2\theta_{u}\sin^2\theta_{d}(m_{u_{2}}-m_{d_{2}})^2\right.\nonumber\\ &+&\left.
\sin^2\theta_{u}\cos^2\theta_{d}(m_{u_{2}}-m_{d_{1}})^2+\cos^2\theta_{u}\sin^2\theta_{d}(m_{u_{1}}-m_{d_{2}})^2\right.\nonumber\\ &+& \left.
\cos^2\theta_{u}\cos^2\theta_{d}(m_{u_{1}}-m_{d_{1}})^2-\sin^2\theta_{u}\cos^2\theta_{u}(m_{u_{2}}-m_{u_{1}})^2\right.\nonumber\\
&-&\left.\sin^2\theta_{d}\cos^2\theta_{d}(m_{d_{2}}-m_{d_{1}})^2\right),
\end{eqnarray}
where $\theta_{u}$ and $\theta_{d}$ are the mixing angles for the
up and down component of the $SU(2)$ doublet respectively, and
$m_{u_{1,2}},~m_{d_{1,2}}$ are the mass eigenvalues. In the case
when all masses are of the same order or we have zero mixing, the
approximation used in Eq.~(\ref{deltaTsimple}) may be applied.

\begin{figure}[htb]
\centerline{ \psfig{figure=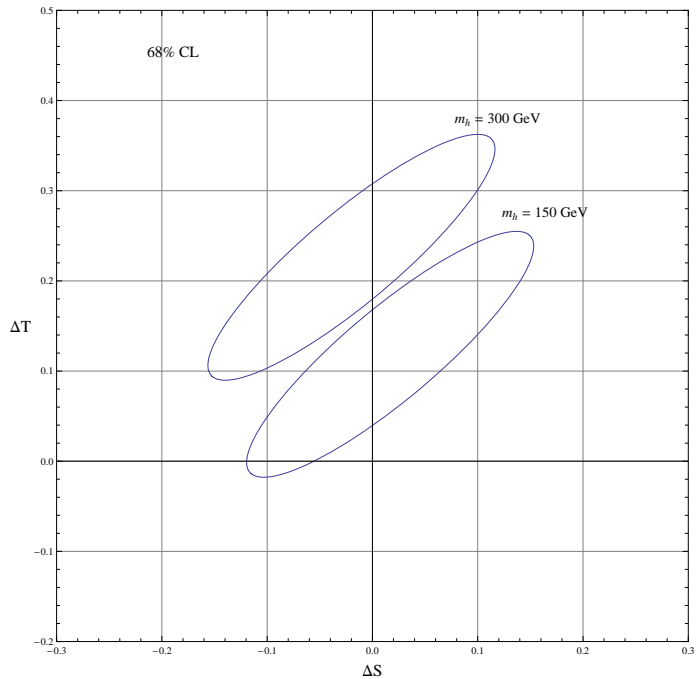,height=10cm}}
\smallskip
\caption{ Values of $\Delta T$ necessary to provide a good fit to electroweak precision data, for two values of the Higgs mass: $m_{h}=150$ GeV and
$m_{h}=300$ GeV.}\label{Fig.Ellipse}
\end{figure}

We present in Fig.~\ref{Fig.Ellipse} the necessary values of new
physics contributions to the $S$ and $T$ parameters, to get
agreement with precision electroweak observables at the
1~$\sigma$~level. The values used here~\footnote{We are thankful
to Jens Erler for providing us with the fit used to generate these
values.} include the recent measurements of $m_W = 80.432 \pm
0.039$~GeV and $m_t = 172.4 \pm 1.2$~GeV, and are consistent with
the ones obtained in Ref.~\cite{Amsler:2008zzb}. The precise
values of $S$ and $T$, however, depend on the observables used in
the fit (see for example Ref.~\cite{:2005ema}). From now on, we
will use the values displayed in Fig.~\ref{Fig.Ellipse} as our
reference values.

The scalars give only a small contribution to the $S$-parameter,
hence the ellipses provide us with information about the values of
$\Delta T$. As can be seen from Fig.~\ref{Fig.Ellipse}, for Higgs
boson masses in the range $m_h \simeq 150$--$300$~GeV, the
necessary value of $\Delta T$ to provide a good fit to the
precision electroweak observables is
\begin{eqnarray}
\Delta T_{150} & \simeq & 0.10 \pm 0.06 \nonumber\\ \Delta T_{300} & \simeq & 0.24 \pm 0.07 \label{deltat150300}
\end{eqnarray}
where $\Delta T_{m_h}$ is the extra contribution to the $T$
parameter necessary to provide a good fit to the precision
electroweak data with a Higgs of mass $m_h(\rm{GeV})$.

Before considering the implications for the sparticles and Higgs
masses one should stress that the enhanced D-term contribution is
obtained from a small deviation of the $\Sigma$ scalar from the
$SU(2)_W$ singlet $u$-direction. The triplet vev leads therefore
to a contribution to the $T$ parameter
\begin{equation}
\Delta T \simeq  \frac{4 \pi g_1^4}{s_W^2 c_W^2 g^4} \frac{m_W^2 u^2}{M_T^4}
\end{equation}
where $M_T$ is the triplet mass. For values of $u$ and the triplet
mass necessary to significantly enhance the tree-level Higgs mass,
these contributions prove to be too small to compensate for the
negative Higgs mass contributions to the $T$-parameter for
$m_{h}\lesssim 300$ GeV. Furthermore, it is clear from the numbers
presented in Eq.~(\ref{deltat150300}) and Fig.~\ref{Fig.Ellipse},
that for SM-like Higgs masses smaller than 150~GeV, the need for
new physics contributions to the precision measurements, and
therefore the constraints on the sparticle spectrum, become
weaker. {}From now on, we will hence concentrate on the region
$m_h \gtrsim 150$~GeV, although somewhat smaller Higgs masses are
still consistent with the presence of enhanced weak $SU(2)$
D-terms.

If sufficiently light, the third generation squark and sleptons,
as well as the non-standard Higgs bosons can lead to large
contributions to the $T$ parameter. The squark contribution is
enhanced by a color factor, but is suppressed by the cancellation
between the D-term and F-term contributions to the left-handed
sbottom-stop splitting. Moreover, in the simplest soft
supersymmetry breaking schemes, the squarks become much heavier
than the sleptons and, from Eq.~(\ref{deltaTsimple}) their
contributions to the $T$-parameter are further suppressed. We
shall first assume that the non-standard Higgs boson mass is a few
times larger than the third generation left-handed slepton masses,
something that is realized, for instance, in the simple scenario
to be discussed in the following section. This means that the
non-standard Higgs bosons will lead to small contributions to the
$T$-parameter compared to the slepton one.

\begin{figure}[htb]
\centerline{ \psfig{figure=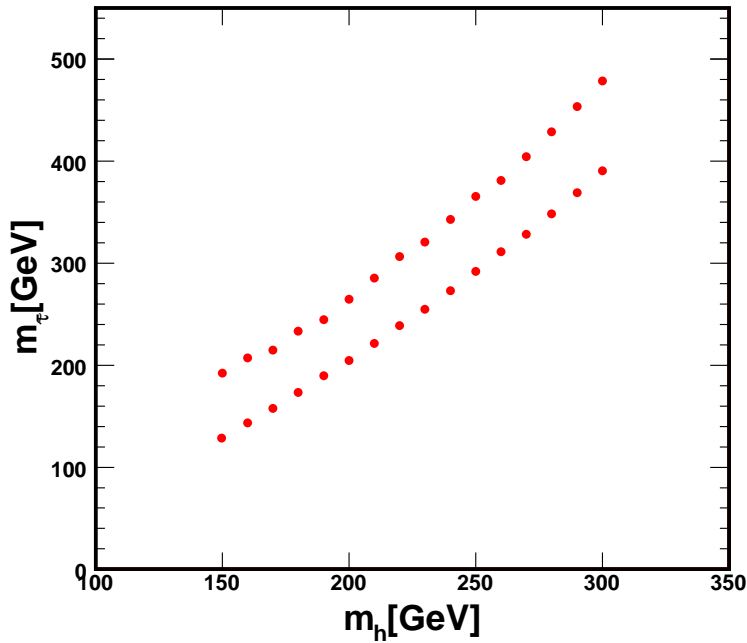,height=10cm}}
\smallskip
\caption{ Range of preferred values of the stau mass [GeV] as a function of the Higgs mass [GeV]. }\label{stau}
\end{figure}

\begin{figure}[htb]
\centerline{ \psfig{figure=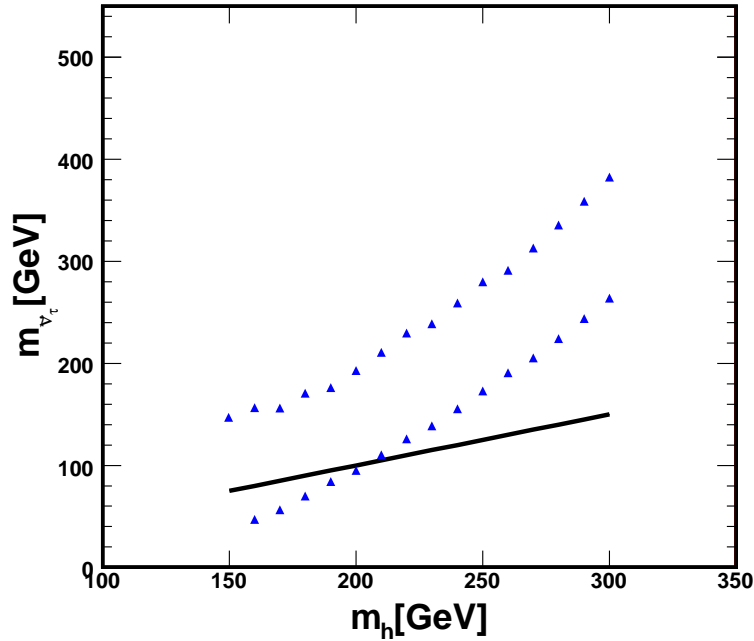,height=10cm}}
\smallskip
\caption{ Range of preferred values of the tau-sneutrino mass
[GeV] as a function of the Higgs mass [GeV]. Sneutrinos below the
black line have masses less than $m_{h}/2$ and therefore are
potential candidates for Higgs decays.}\label{sneutrino}
\end{figure}

\begin{figure}[htb]
\centerline{ \psfig{figure=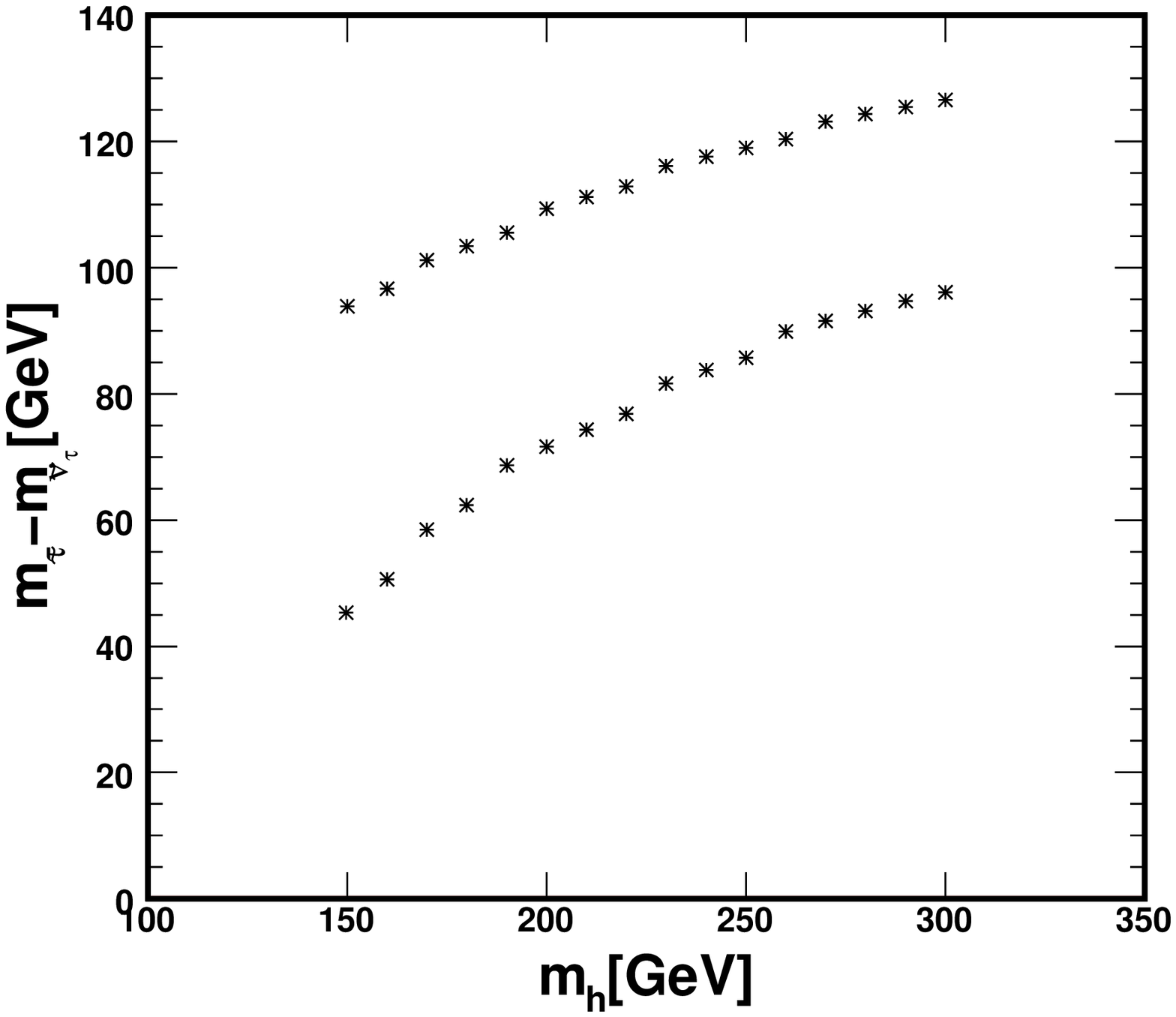,height=10cm}}
\smallskip
\caption{ Range of preferred values of the stau--sneutrino mass difference [GeV] as a function of the Higgs mass [GeV]. }\label{difference}
\end{figure}

Considering that only the sleptons contribute to the $T$
parameter, the requirement of restoring agreement with precision
electroweak observables defines the allowed range of slepton
masses. Using the values obtained in Fig.~\ref{Fig.Ellipse} for a
Higgs mass of 150~GeV, ignoring mixing in the squark and slepton
sectors and assuming that the Higgs mass receives two-loop
contributions from third generation squarks with masses of
about~$\sim~1$~TeV, this leads, for the lower (upper) required
values of the new physics contribution to the $T$ parameter, to
sneutrino and stau masses:
\begin{eqnarray}
m_{\tilde \nu} & \simeq & 150 \;\; (30)\; {\rm GeV} \nonumber\\ m_{\tilde \tau} & \simeq & 195 \;\; (120) \; {\rm GeV.}
\end{eqnarray}
For a Higgs mass of 300~GeV, instead, we get a relatively heavier spectrum
\begin{eqnarray}
m_{\tilde \nu} & \simeq & 380 \;\; (260)\; {\rm GeV} \nonumber\\ m_{\tilde \tau} & \simeq & 480 \;\; (395) \; {\rm GeV.}
\end{eqnarray}
The sleptons should then acquire masses of the order of a few
hundred GeV, with a relatively large mass splitting, which grows
for larger Higgs masses and is in the range of a few tens of GeV
to more than a 100 GeV. The results are shown in Figs.~\ref{stau},
\ref{sneutrino} and \ref{difference}. In Fig.~\ref{sneutrino} we
have also drawn a straight line with slope of $m_{h}/2$.
Sneutrinos below the line are potential candidates for Higgs
decays.


Furthermore, there exists a lower bound on the sneutrino mass,
$m_{\tilde{\nu}_{\tau}}\gtrsim 40$ GeV, coming from the invisible
decay width of the $Z$ gauge boson measured at LEP\cite{:2005ema}.
This bound, as shown in Fig.~\ref{sneutrino}, is almost
automatically fulfilled for sneutrino masses in the range selected
by precision electroweak measurements for $m_h > 150$~GeV included
in the plots.

\section{Phenomenological Properties}
\subsection{Higgs Boson Searches}

One of the most important differences of this model with the
standard implementation of the MSSM is the Higgs mass range, which
has an important impact on its searches. For the mass range
150~GeV~$\simlt m_h \simlt$~300~GeV that we are considering in
this article, the most important SM-like Higgs boson search
channels will be its decays into charged and neutral gauge bosons,
namely
\begin{eqnarray}
H & \to & W^{\pm} \; W^{\mp} \nonumber\\ H & \to & Z \; Z.
\end{eqnarray}

These searches will become very efficient in the first years of
the LHC run~\cite{HsearchesLHC}, and therefore we expect these
kind of models to be probed with the first few fb~$^{-1}$ of LHC
running. Moreover, if the Higgs mass is below 190~GeV, then even
the Tevatron will be able to probe these models in the coming
years~\cite{HsearchesTeV}. Indeed, the Tevatron has already
excluded the existence of a SM-like Higgs boson with mass between
160 and 170~GeV at the 95\% confidence level, and is expected to
probe the whole range of masses, $m_{h}\in [150,190]$~GeV, by the
end of next year.

The Higgs boson searches would change if there would be new,
supersymmetric decays of the Higgs bosons. If we assume that the
lightest superpartner of the SM particles is the light
tau-sneutrino discussed in the last section, there is only a small
region of parameters where these decays would be open, and only
for masses below 190~GeV, that coincides with the range to be
explored by the Tevatron collider. As can be seen from
Fig.~\ref{sneutrino}, for larger values of the Higgs mass, the
tau-sneutrino is sufficiently heavy as to avoid any Higgs decay
into light supersymmetric particles.

For Higgs boson masses below 190~GeV the sneutrinos may be light
enough to allow on-shell decays of the Higgs into two sneutrinos.
Moreover, the coupling of third generation sneutrinos to the Higgs
boson is enhanced with respect to the one in the MSSM, due to the
enhanced $SU(2)$ D-term contribution and is given by
\begin{equation}
g_{h \tilde{\nu}_\tau \tilde{\nu}_\tau} \simeq -i \; \frac{(g^2 \Delta + g_Y^2) \; v}{2 \sqrt{2}}.
\end{equation}
Due to this enhanced coupling, the Higgs boson may have a
significant decay branching ratio into light sneutrinos, therefore
avoiding the Tevatron bounds. More quantitatively, assuming masses
above the 2~$m_W$ threshold, the decay width for decays into two
vector bosons is given by
\begin{equation}
\Gamma(h \to VV) \simeq \frac{G_F (|Q_V| + 1)\; m_h^3}{\sqrt{2}\; 16\; \pi} \left( 1 - \frac{4 m_V^2}{m_h^2} + \frac{ 12 m_V^4}{m_h^4} \right) \left( 1- \frac{4 m_V^2}{m_h^2} \right)^{1/2}
\end{equation}
where $G_F$ is the Fermi constant, $Q_V$ is the charge of the
massive gauge boson, $V = W^{\pm},\; Z$, and $M_V$ is its mass.
For the decay into sneutrinos, instead, the result is
\begin{equation}
\Gamma(h \to \tilde{\nu}_{\tau} \tilde{\nu}_{\tau}) \simeq \frac{(g^2 \Delta + g_Y^2)^2 \; v^2  }{128 \; \pi \; m_h} \left( 1 - \frac{4
m_{\tilde{\nu}_{\tau}}^2}{m_h^2} \right)^{1/2} .
\end{equation}

For instance, for a Higgs mass $m_h \simeq 165$~GeV and a
sneutrino mass of about 70~GeV, the branching ratio of the Higgs
into a pair of $W^{\pm}$ gauge bosons is reduced to less than half
of its SM value, avoiding the current Tevatron bound. Somewhat
smaller or larger values of the sneutrino masses, within the range
consistent with precision measurements and smaller than a half of
the Higgs mass, lead to more or less restrictive Tevatron bounds
on this model, respectively.

\subsection{Supersymmetric Particle Spectrum}

In this section, we will work in the moderate $\tan\beta$ limit,
which for universal scalar masses, tend to lead to heavy
non-standard Higgs bosons and is therefore close to the so-called
decoupling limit in the Higgs sector. In the foregoing discussion
we assumed that the mixing of left-handed and right-handed third
generation sleptons is small. Wether the lightest stau is mostly
left-handed or right-handed depends on the supersymmetry breaking
mechanism. We will assume that $\tilde{\tau}_{1}$ is mostly
left-handed while $\tilde{\tau}_{2}$ is mostly right-handed, and
present a mechanism that realizes this possibility below.

The tree-level third generation Yukawa couplings are given by
\begin{equation}
h_{t}=\frac{gm_{t}}{\sqrt{2}m_{W}\sin\beta},\hspace{1.0cm} h_{b}=\frac{gm_{b}}{\sqrt{2}m_{W}\cos\beta},\hspace{1.0cm}
h_{\tau}=\frac{gm_{\tau}}{\sqrt{2}m_{W}\cos\beta}. \label{tanbeta}
\end{equation}
As is well known, $h_{b}$ and $h_{\tau}$ can become comparable to
$h_{t}$ in the large $\tan\beta$ regime. As previously mentioned,
we are interested in the moderate $\tan\beta$ limit and therefore
only the top quark Yukawa will be of importance. In the examples
found later in the text, we take the reference value of
$\tan\beta=10$.

We shall assume that Yukawa couplings for the first and second
generation quarks and leptons are generated by adding a massive
Higgs-like pair of doublets which transform under $SU(2)_{2}$, as
was suggested in~\cite{Batra:2003nj}, and work with this minimal
spectrum. With this minimal spectrum, the $\beta_i$ function
coefficients are given by
\begin{equation}
b_Y = \frac{36}{5}, \;\;\;\;\;\;\;\;\;\;\;\;\;\; b_{22} = 1,  \;\;\;\;\;\;\;\;\;\;\;\;\;\; b_{21} = -1  \;\;\;\;\;\;\;\;\;\;\;\;\;\; b_3 = -3,
\end{equation}
where $b_{22}$ and $b_{21}$ correspond to the $SU(2)_2$ and
$SU(2)_1$ gauge groups, while $b_Y$ and $b_3$ correspond to the
hypercharge and strong interactions, respectively. The above
values should be compared to the MSSM case:
\begin{equation}
b_Y = \frac{33}{5}, \;\;\;\;\;\;\;\;\;\;\;\;\;\; b_{2} = 1,  \;\;\;\;\;\;\;\;\;\;\;\;\;\; b_{3} = -3.
\end{equation}
This will generally upset unification of
couplings~\cite{Maloney:2004rc}. Let us stress, however, that for
the chosen low energy values of these couplings, the $SU(2)_2$
coupling $\alpha_2$ becomes close to $\alpha_Y$ and $\alpha_3$ at
scales of about the standard grand unification scale,
$M_{GUT}\approx 2\times 10^{16}$ GeV.

The modified $\beta$ function coefficients also have consequences
for the soft supersymmetry breaking parameters. For instance,
assuming a common value for the gaugino mass, $M_{1/2}$, at the
messenger scale, the running of the Bino mass is modified, making
the lightest neutralino lighter than in the MSSM for the same
value of $M_{1/2}$. Regarding the third generation masses, if we
assume a common gaugino mass, they will be pushed to large values
driven by the strong $SU(2)_1$ gaugino effects. Therefore, for the
third generation sleptons to remain light, we need a mechanism of
SUSY breaking that keeps the third generation soft scalar masses
small at low energies while the strong $SU(2)_{1}$ coupling
$g_{1}$ runs to large values.

We can accomplish this with a model where SUSY breaking is
transmitted to the visible sector via only the $SU(3)_c \times
SU(2)_{2} \times U(1)_Y$ gauginos. The $SU(2)_{1}$ gauge sector
then does not contribute to the running of soft third generation
masses. For simplicity, the input parameters for the model are:
$\tan \beta$; a universal gaugino mass $M_{1/2}$ for the
$SU(3)_c$, $SU(2)_2$ and $U(1)_Y$ gauginos; a universal scalar
mass $m_{0}$ for all squarks and sleptons; soft-supersymmetry
breaking parameters $m_{H_{u}}$ and $m_{H_{d}}$ for the two Higgs
bosons which interact under $SU(2)_1$; and we assume positive
sign($\mu$) and $A_{t}=A_{b}=A_{\tau}=0$ at the messenger scale
$M$ which is of the order of $M_{GUT}$. The $SU(2)_1$ gaugino
remains massless, but acquires a mass of order $g_1 u$, together
with the $SU(2)_1$ gauge bosons, after the breakdown of $SU(2)_1
\times SU(2)_2 \to SU(2)_W$.

Lets take a look at the one-loop renormalization group equations
(RGE) for the left- and right-handed soft stau masses at energy
scales larger than the gauge symmetry breaking scale
$SU(2)_{1}\times SU(2)_{2}\rightarrow SU(2)_{W}$,
\begin{eqnarray}
16\pi^2\frac{d}{dt}m^{2}_{L_{3}}&=&-\frac{6}{5}g^{2}_{Y}|M_{Y}|^2 -\frac{3}{5}g^{2}_{Y}S\label{RGE1}\\
16\pi^2\frac{d}{dt}m^{2}_{\tilde{\tau}_{R}}&=&-\frac{24}{5}g^{2}_{Y}|M_{Y}|^2 +\frac{6}{5}g^{2}_{Y}S\label{RGE2}
\end{eqnarray}
where $S$ is given by,
\begin{equation}
S\equiv\rm{Tr}[Y_{i}m^2_{\phi_{i}}]=m^2_{H_{u}}-m^{2}_{H_{d}}+\rm{Tr}[m^2_{Q}-m^2_{L}-2m^2_{\bar{u}}+m^{2}_{\bar{d}}+m^{2}_{\bar{e}}].
\end{equation}
In Eqs.(\ref{RGE1}) and (\ref{RGE2}) we have neglected the terms
proportional to the $\tau$-Yukawa coupling, which remain small
unless $\tan\beta$ is very large and, as stressed above, we have
assumed that our symmetry breaking scheme is $SU(2)_{1}$ blind.
Furthermore we take $m_{H_{u}}=m_{H_{d}}$ and since we have chosen
a universal scalar mass $m_{0}$ at high energy, this implies that
$S$ vanishes and does not run with energy. Therefore, given that
the $U(1)_{Y}$ charges for left-handed sleptons are smaller than
the $U(1)_{Y}$ charges of the right-handed sleptons, this
naturally leads to the left-handed stau soft masses being smaller
than the right-handed ones, as can be seen from Eqs.(\ref{RGE1})
and (\ref{RGE2}). This is in contrast to what happens in the
MSSM~\cite{Ellis:2002iu,Covi:2007xj,Ellis:2008as}, where the
non-universal requirement $m_{H_{d}}-m_{H_{u}}>0$ is necessary to
obtain smaller left-handed masses for the third generation
sleptons. In the case of soft gaugino masses, their one-loop RGE
take the form,
\begin{equation}
16\pi^2\frac{d}{dt}M^{2}_{i}=4b_{i} \;g_i^2 \; M_i^2\label{RGE3}
\end{equation}
where, as stressed above, up to the $SU(2)_{1}\times SU(2)_{2}$
breaking scale $b_{i}=(36/5,-1,1,-3)$ for the $U(1)_Y$, $SU(2)_1$,
$SU(2)_2$ and $SU(3)_c$ gauge groups, and after the breakdown of
$SU(2)_{1}\times SU(2)_{2}\rightarrow SU(2)_{W}$, we get the
standard MSSM result, $b_{i}=(33/5,1,-3)$.

\begin{figure}[htbp]

    \begin{minipage}[b]{0.5\linewidth}

        \centering
        \includegraphics[scale=0.75]{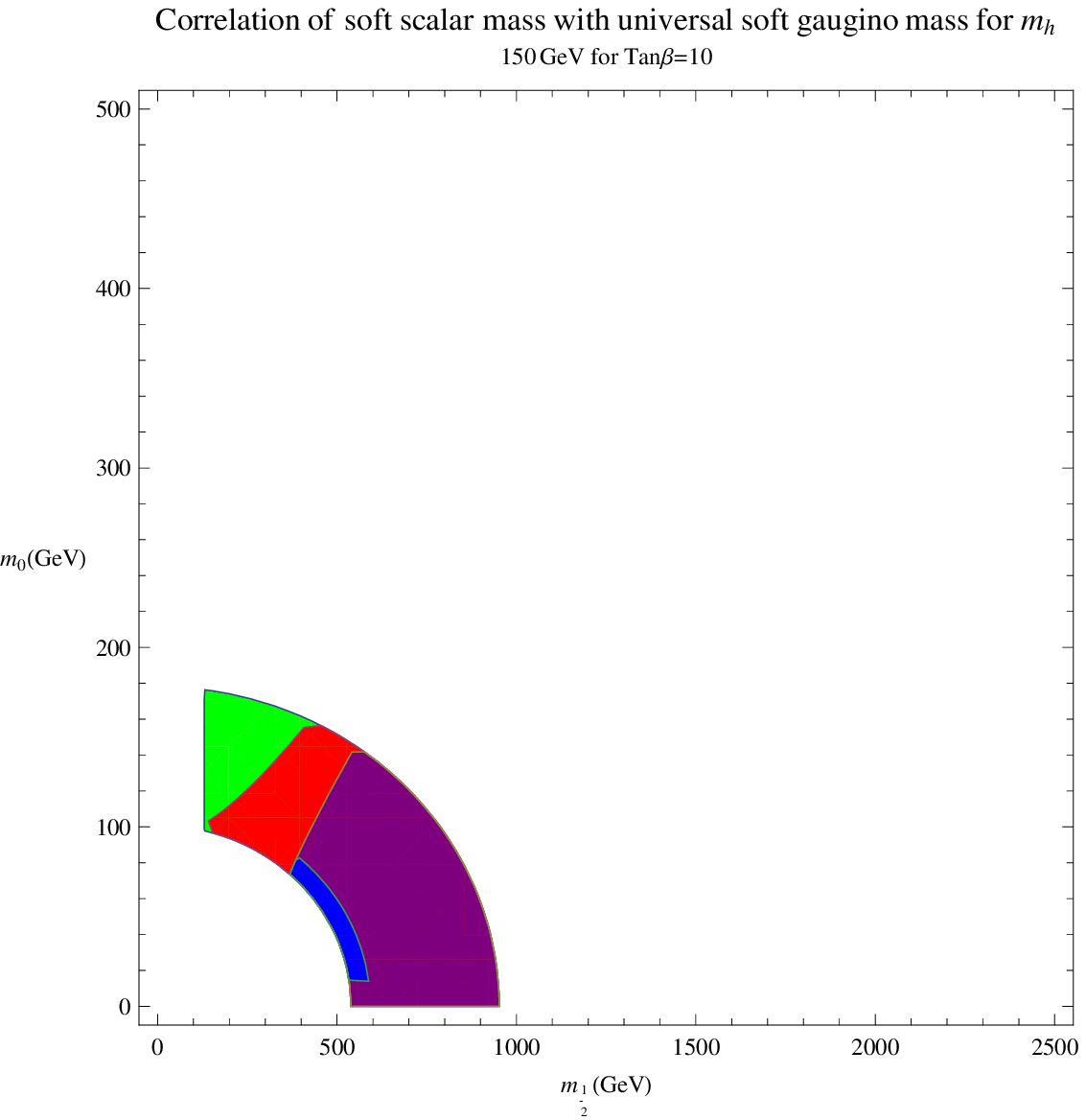}
        \label{collfig23}

    \end{minipage}
    \hspace{0.2cm}
    \begin{minipage}[b]{0.5\linewidth}

        \centering
        \includegraphics[scale=0.75]{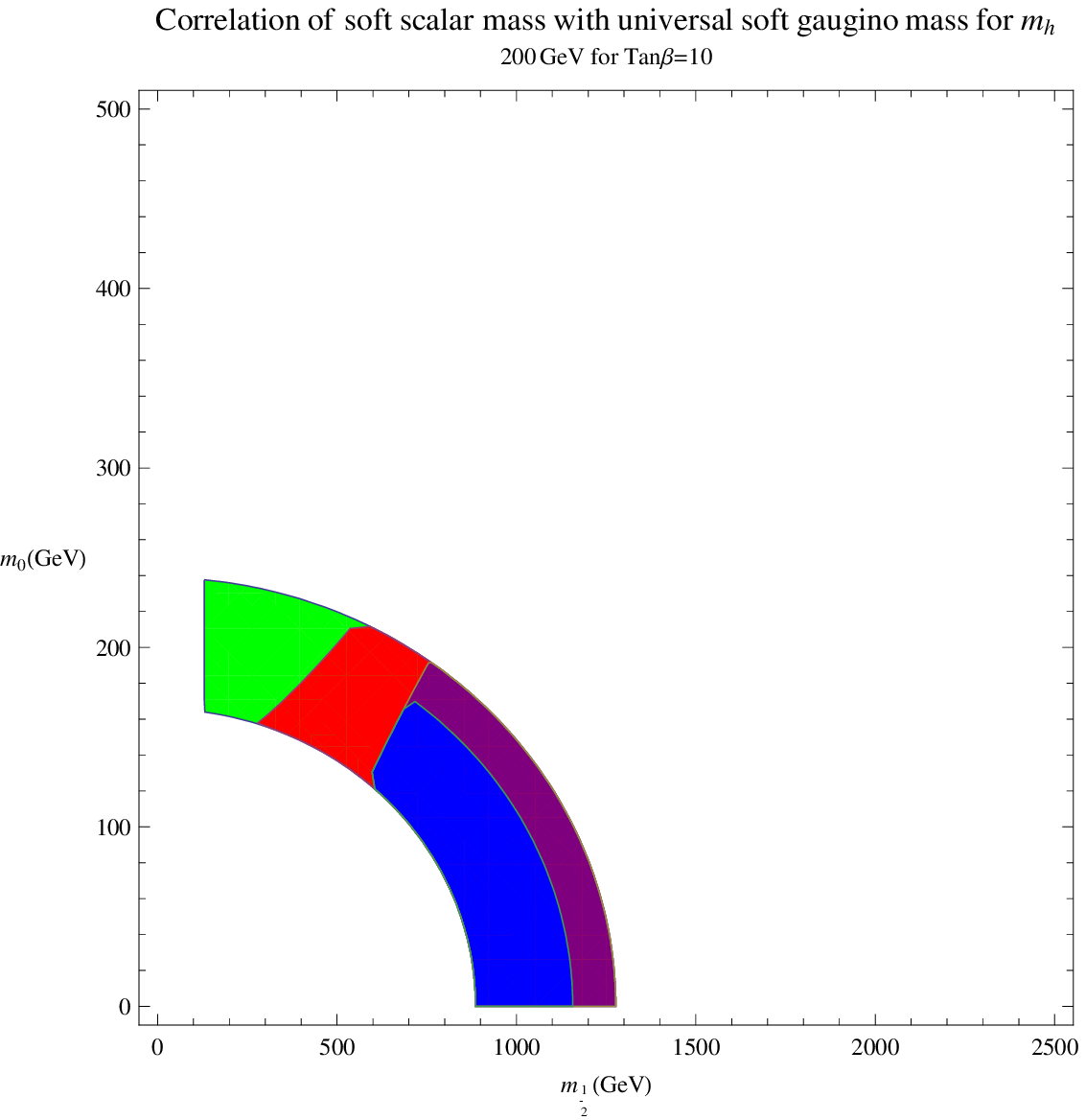}
        \label{collfig24}

    \end{minipage}
    \vspace{0.0cm}

    \begin{minipage}[b]{0.5\linewidth}

        \centering
        \includegraphics[scale=0.75]{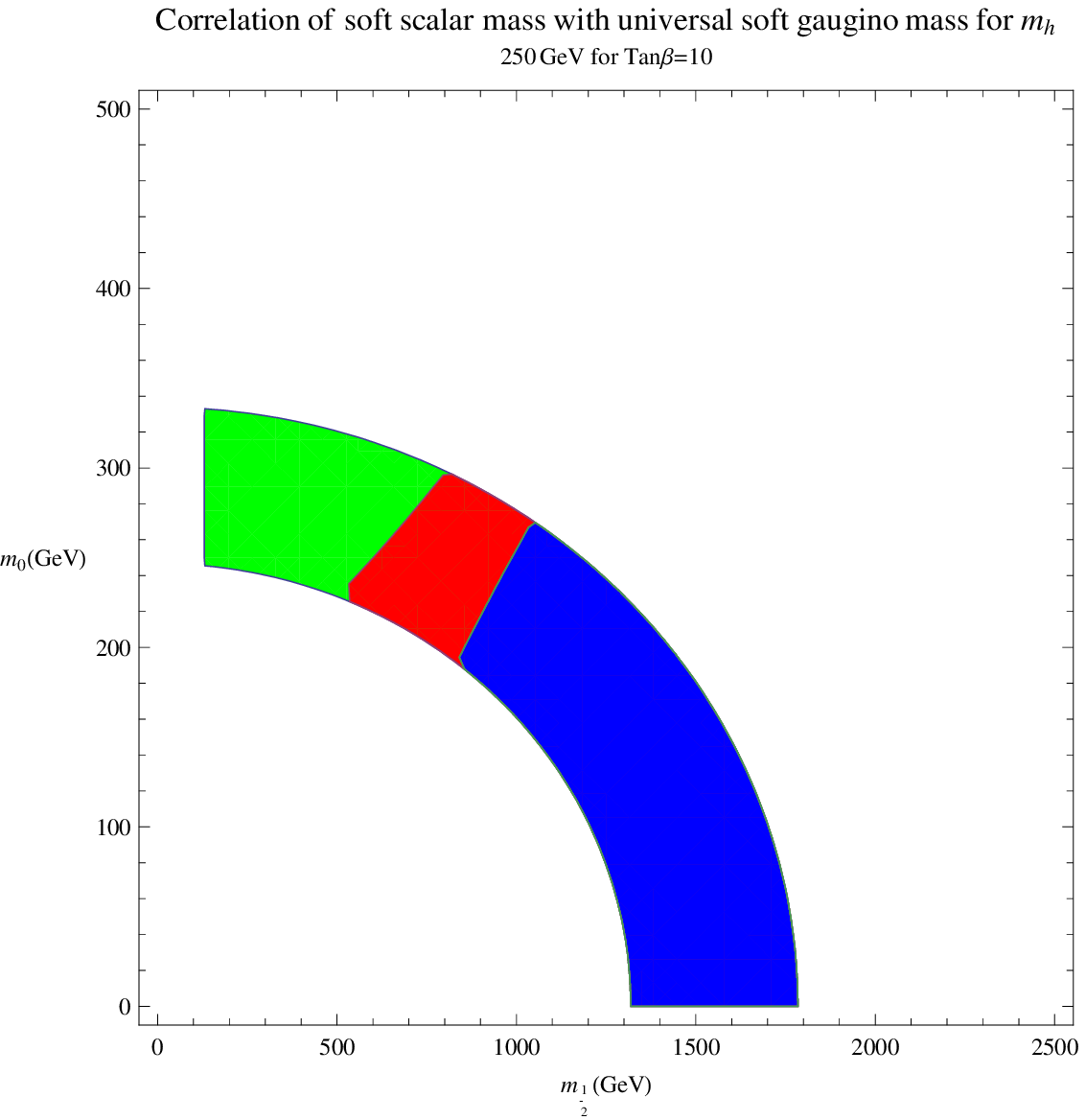}
        \label{collfig25}

    \end{minipage}
    \hspace{0.2cm}
    \begin{minipage}[b]{0.5\linewidth}

        \centering
        \includegraphics[scale=0.75]{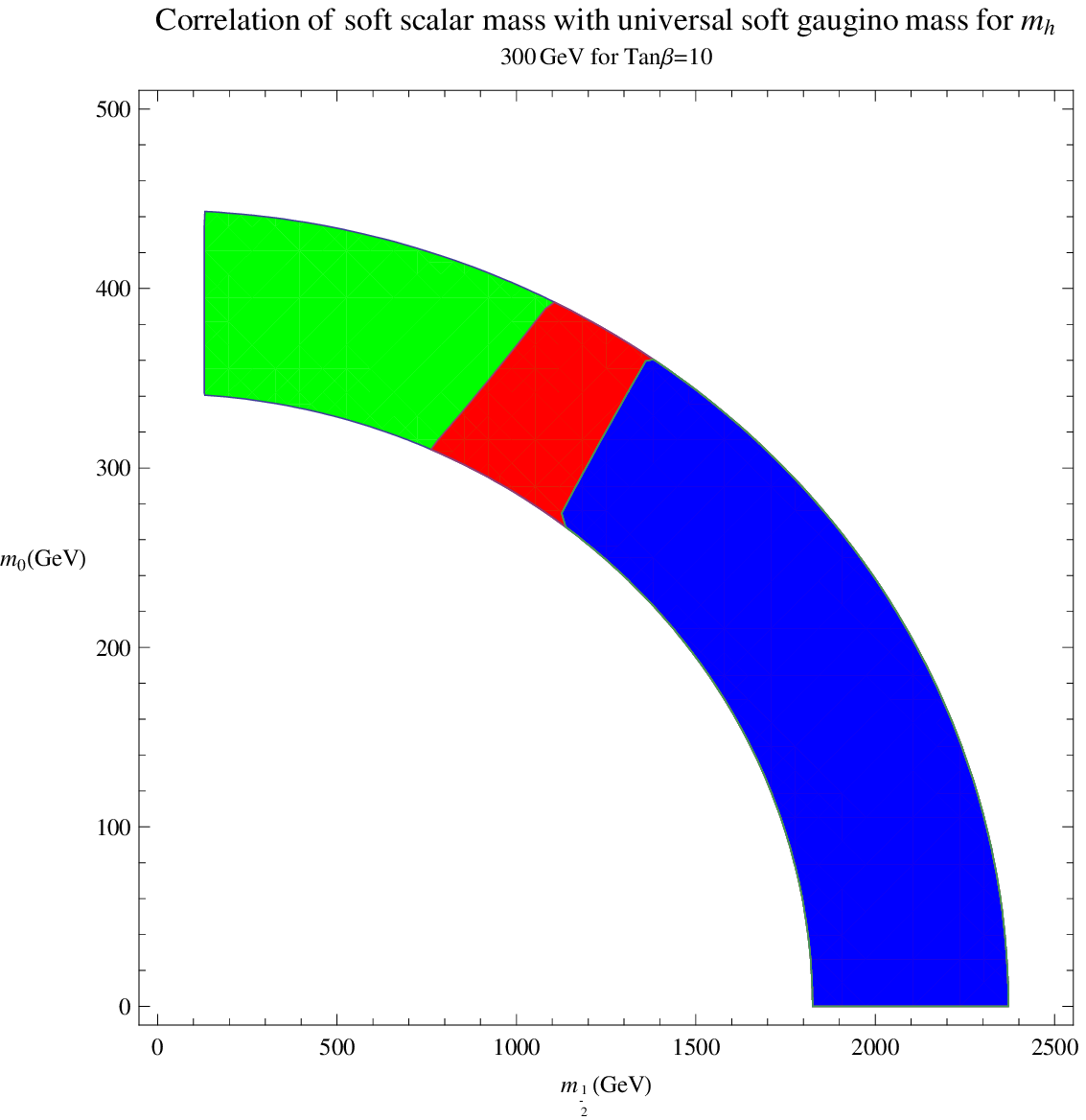}
        \label{collfig26}

    \end{minipage}
\caption{Correlation of $m_0$ and $M_{1/2}$ for different Higgs
masses and $\tan\beta=10$. Red represents the region where
$\tilde{\nu}_{\tau}$ is the NLSP and
$m_{\tilde{\tau}_{1}}>m_{\tilde{\chi}_{0}}$, blue and purple are
the regions where $\tilde{\nu}_{\tau}$ is the NLSP and
$m_{\tilde{\chi}_{0}}>m_{\tilde{\tau}_{1}}$ and staus decay
predominantly to on shell Ws and tau sneutrinos (blue region) or
to off-shell Ws and sneutrinos or off-shell binos and taus (purple
region). Green is the region where $\chi_{0}$ is the
NLSP.}\label{m0m1/2}
\end{figure}

The solution of the RG equations show that the soft supersymmetry
breaking parameters $m_{L_3}^2$ and $m_{\tilde{\tau}_{R}}$ are
approximately given by
\begin{equation}
\label{ml3} m_{L_3}^2 \simeq m_0^2 + 0.04 \; M_{1/2}^2, \quad\quad m^2_{\tilde{\tau}_{R}}\simeq m_0^2+0.15 \; M_{1/2}^2,
\end{equation}
while the low energy Bino and Wino masses are given by
\begin{equation}
M_Y \simeq 0.35 \; M_{1/2}, \quad\quad M_{2}\simeq 0.8\; M_{1/2}^2.
\end{equation}
The requirement of getting the appropriate stau and tau-sneutrino
masses for a given value of the Higgs mass and $\tan\beta$ defines
the allowed parameter space in the $m_0$--$M_{1/2}$ plane. In
particular, looking at Eq.(\ref{deltaTsimple}) and (\ref{ml3}), we
notice that for a fixed Higgs mass and in the absence of stau
mixing there is an ellipsoidal area in the $M_{1/2}$--$m_{0}$
plane where our spectrum fits electroweak precision data at a 68
\% confidence level. In Fig.~\ref{m0m1/2} we plot this correlation
for several Higgs masses neglecting the stau mixing. For large
values of $|\mu|$, there is a lower bound of $M_{1/2}\gtrsim 130$
GeV coming from the LEP bound on chargino masses~\cite{:chargino}.
We see that the preferred region of parameter space is relatively
thin and moves to higher soft masses as the Higgs mass increases,
roughly maintaining its thickness. The red (medium light gray),
blue (dark gray) and purple (medium dark gray) regions in the
figure have the additional constraint of a $\tilde{\nu}_{\tau}$
NLSP, whereas in the green (light gray) region
$m_{\tilde{\chi}_{0}}<m_{\tilde{\nu}_{\tau}}$. The blue and purple
regions instead, have the constraint that the bino is heavier than
the stau, $m_{\tilde{\tau}} < m_{\tilde{\chi}_0}$. Finally, in the
blue region the stau dominant decay mode is into an on-shell
$W^{\pm}$ and a tau-sneutrino; this does not happen in the purple
region, where it decays off-shell to $W^{\pm}$ and a tau-sneutrino
and also into an off-shell Bino and a $\tau$ lepton. We can see
from the figure that, as expected, the region where
$\tilde{\nu}_{\tau}$ is the NLSP becomes more prominent for
smaller values of $m_{0}$. Furthermore, for larger values of the
Higgs mass, the mass difference between the stau and the
tau-sneutrino is sufficiently large to allow on-shell decays into
a sneutrino and a charged gauge boson in most of the parameter
space.
We will assume in this work, as is natural if the messenger scale
is somewhat smaller than the GUT scale, that the LSP is the
gravitino and, as can be seen in Fig.~\ref{m0m1/2} appropriate
boundary conditions at high energy can be chosen such that the
NLSP is the lightest tau sneutrino, $\tilde{\nu}_{\tau}$. This
should be compared with, for instance, the standard low scale
gaugino mediated scenario, in which the right-handed staus are
naturally the
NLSP~\cite{Schmaltz:2000ei,Schmaltz:2000gy,Medina:2006hi}.

Due to our mechanism of SUSY breaking and the new big D-term
contributions to third generation sleptons, there is a lower bound
on the soft scalar universal mass coming from the requirement of
the sneutrino to be in the right mass range to lead to agreement
with precision electroweak observables. This can be seen in
Fig.~\ref{m0m1/2}, where for example, for a Higgs mass $m_h \simeq
150$~GeV, $m_{0}\gtrsim 40$ GeV for $M_{1/2}=500$ GeV to fulfill
this requirement.

In the case of a universal gaugino mass, $M_{1/2}$, at high
energy, we naturally obtain that the lightest neutralino,
$\tilde{\chi}^{0}_{1}$, is mostly bino-like, while the second
lightest neutralino, $\tilde{\chi}^{0}_{2}$, is mostly wino-like.
This implies that right-handed squarks mostly decay to
$q~\tilde{\chi}^{0}_{1}$. From Fig.~\ref{m0m1/2}, depending on the
initial masses, $m_{0}$ and $M_{1/2}$, we see that we can have
either $m_{\tilde{\tau}_{1}}<m_{\tilde{\chi}^{0}_{1}}$ or
$m_{\tilde{\chi}^{0}_{1}}<m_{\tilde{\tau}_{1}}$ for a small scalar
mass $m_{0}$ and a moderate gaugino mass $M_{1/2}$, once we take
into account the D-term corrections. We also expect the
left-handed first and second family squarks and sleptons to be
heavier than the lightest neutralino since their masses run with
$SU(3)_c$ and $SU(2)_{2}$ couplings respectively. With these
considerations in mind, for a light sneutrino NLSP, we infer the
following possible mass spectra:
\begin{eqnarray}
m_{\tilde{G}}&<&m_{\tilde{\nu}_{\tau}}<m_{\tilde{\tau}_{1}}< m_{\tilde{\chi}^{0}_{1}}< m_{\tilde{\chi}^0_2}, m_{\tilde{\chi}_1^{\pm}} <
m_{\tilde{\tau}_{2}},m_{\tilde{l}_{R}},m_{\tilde{l}_{L}}...\\ m_{\tilde{G}}&<&m_{\tilde{\nu}_{\tau}}<m_{\tilde{\chi}^{0}_{1}}<m_{\tilde{\tau}_{1}} <
m_{\tilde{\chi}^0_2}, m_{\tilde{\chi}_1^{\pm}} <m_{\tilde{\tau}_{2}},m_{\tilde{l}_{R}},m_{\tilde{l}_{L}}...
\end{eqnarray}
which are realized in Tables~\ref{colltable1}, \ref{colltable2}
and \ref{colltable3}.

In order to calculate the particle spectrum at low energies for
some sample points at high energies and to calculate the main
decays we used SDECAY~\cite{Muhlleitner:2003vg}. SDECAY works
mainly within the MSSM and thus we had to slightly modify the RGE
used in the program to make it suitable for our model. From the RG
running we expect modifications to the spectrum at high energies.
Furthermore SDECAY only calculates two body decays of sparticles.
To understand the behavior at low energies, let us write the tree
level stau mass matrix, taking into account the non-zero trilinear
terms and the left-right mixing,
\begin{equation}
m^2_{\tilde{\tau}}=\begin{array}{cc}
\begin{pmatrix}
m^2_{L_{3}}+\Delta_{L_{3}}-\frac{g^2}{4}(\Delta-1) v^2 \cos 2\beta & m_{\tau}(A_{\tau}-\mu\tan\beta) \\
 m_{\tau}(A_{\tau}-\mu\tan\beta) & m^2_{\tilde{\tau}_{R}}+\Delta e_{3}
\end{pmatrix}
\end{array}\label{staumatrix}
\end{equation}
where $m_{L_{3}}$ and $m_{\tilde{\tau}_{R}}$ are the third
generation left- and right-handed slepton soft breaking masses
respectively, $\Delta_{L_{3}}=(-1/2+\sin^2\theta_{w})\cos 2\beta\;
m^2_{Z}$ and $\Delta_{e_{3}}=\sin^2\theta_{w}\cos 2\beta\;
m^2_{Z}$. We can see that the mass eigenstates will have an
admixture of left- and right-handed components, mainly due to the
mixing caused by the $\mu$ term in Eq.(\ref{staumatrix}). However,
for $\tan\beta \simeq 10$, due to the smallness of the $\tau$
mass, we expect this admixture to be small. Similarly, the tau
sneutrino modified mass is given by,
\begin{equation}
m^2_{\tilde{\nu}_{\tau}}=m^2_{L{3}}+\Delta\nu_{\tau}+\frac{g^2}{4}(\Delta-1) v^2\cos 2\beta
\end{equation}
where $\Delta_{\nu_{\tau}}=\cos 2\beta\; m^2_{Z}/2$.

Furthermore, minimizing the tree level Higgs potential we find the
following expression for $\mu$,
\begin{equation}
\mu^2=\frac{1}{2}\left(\frac{m^2_{H_{u}}-m^2_{H_{d}}}{\cos2\beta}-m^2_{H_{u}}-m^2_{H{d}}-m_\Delta^2\right)
\end{equation}
where $m_\Delta^2=(g^2_{1}+ g^2_{2}\Delta)v^2/2 $, and likewise the tree-level Higgs masses are given by,
\begin{eqnarray}
m^2_{A}&=& \frac{m^2_{H_{u}}-m^2_{H_{d}}}{\cos2\beta}-m_\Delta^2,\label{mA}\\
m^2_{h^{0}}&=&\frac{1}{2}\left(m_{A}^2+m_\Delta^2-\sqrt{(m_{A}^2-m_\Delta^2)^2+4m_\Delta^2 m_{A}^2\sin 2\beta}\right),\label{h^0}\\
m^2_{H^{0}}&=&\frac{1}{2}\left(m_{A}^2+m_\Delta^2+\sqrt{(m_{A}^2-m_\Delta^2)^2+4m_\Delta  ^2 m_{A}^2\sin 2\beta}\right),\\
m^2_{H^{+}}&=&m^2_{A}+\frac{g^2}{2}\Delta v^2.\label{MHplus}
\end{eqnarray}
Observe that for $\Delta = 1$, the above expressions reduce to the well known MSSM ones.

\subsection{Collider Signatures}

The particle spectrum and the most important branching ratios for
some sample points are presented in
Tables~\ref{colltable1},~\ref{colltable2} and~\ref{colltable3}.
Table~\ref{colltable1} presents a case of sneutrinos which are
light enough to allow the decay of a Higgs into two on-shell
tau-sneutrinos. Similar to the example discussed in the previous
section, the Higgs decay branching ratio into a pair of
$W^{\pm}$'s is reduced to a value that is less than half of the SM
result, therefore avoiding the current Tevatron bounds.

As can be seen from Table~\ref{colltable2} and~\ref{colltable3},
two body decays, $\tilde{\tau}^{\pm}_{1}\rightarrow
W^{\pm}\tilde{\nu}_{\tau}$, tend to be kinematically allowed and
thus may be an important decay modes for the light staus.
Therefore, we expect relatively hard leptons and$/$or jets plus
missing energy to be the an important signature from
$\tilde{\tau}_{1}$ decays. This is very different from regular
gaugino mediated SUSY breaking scenarios in the MSSM where the
lightest stau tends to decay into $\tau$'s and gravitinos, and the
heaviest stau, since the mass difference,
$m_{\tilde{\tau}}-m_{\tilde{\nu}_{\tau}}$ is small, tends to decay
into an on-shell $\tau$ and a Bino. Indeed, the small
sneutrino-stau mass difference allows only 3-body decays, which
through an off-shell $W$, leads to soft fermions: $f\bar{f}'$ plus
missing energy~\cite{Covi:2007xj}.

\begin{table}[htbp]
\begin{center}
\begin{tabular}{|c|c|c|}
\hline Sparticle & Mass[GeV] & Dominant decay modes\\
\hline $\tilde{g}$ & 1146 & $\tilde{q}_{L}q$ (18)$\%$,\hspace{1 mm}$\tilde{q}_{R}q$ (36)$\%$,\hspace{1 mm}$\tilde{b}_{1,2}b$ (23) $\%$,\hspace{1 mm} $\tilde{t}_{1}t$ (23) $\%$ \\
$\tilde{u}_{L},\tilde{d}_{L}$ & 1047, 1050 & $\tilde{\chi}^0_{2}q$ (32) $\%$, \hspace{1 mm}$\tilde{\chi}^{\pm}_{1}q'$ (64) $\%$\\ $\tilde{u}_{R},\tilde{d}_{R}$ & 1010,1007 & $\tilde{\chi}^0_{1}q$ (99) $\%$ \\
$\tilde{t}_{1}$ & 804& $\tilde{\chi}^+_{1}b$ (34) $\%$,\hspace{1 mm}$\tilde{\chi}^0_{1}t$ (27) $\%$,\hspace{1 mm}$\tilde{\chi}^0_{2}t$ (14) $\%$\\
$H^{+}$ & 724& \\
$A$     & 708 &\\
$\tilde{\chi}^{0}_{4}$ & 644& $\tilde{\chi}^{\pm}_{1}W^{\mp}$ (54) $\%$,\hspace{1 mm}$\tilde{\chi}^0_{2}h$ (17) $\%$\\
$\tilde{\chi}^{\pm}_{2}$ & 644& $\tilde{\chi}^0_{2}W^{\pm}$ (28) $\%$,\hspace{1 mm}$\tilde{\chi}^{\pm}_{1}Z$ (26)
$\%$,\hspace{1 mm}$\tilde{\chi}^{\pm}_{1}h$ (18) $\%$\\
$\tilde{\chi}^0_{3}$ & 630& $\tilde{\chi}^{\pm}_{1}W^{\mp}$ (57) $\%$,\hspace{1 mm}$\tilde{\chi}^0_{2}Z$ (25) $\%$\\
$\tilde{\chi}^0_{2}$ & 385& $\tilde{\nu}_{\tau}\nu_{\tau}$ (48) $\%$,\hspace{1 mm}$\tilde{\tau}^{\pm}_{1}\tau^{\mp}$ (41) $\%$\\ $\tilde{\chi}^{\pm}_{1}$ & 385& $\tilde{\nu}_{\tau}\tau^{\pm}$ (50) $\%$,\hspace{1 mm}$\tilde{\tau}^{\pm}_{1}\nu_{\tau}$ (38) $\%$\\
$\tilde{e}_{L}$ & 346& $\tilde{\chi}^0_{1}e$ (100) $\%$\\
$\tilde{\nu}_{e}$ & 337& $\tilde{\chi}^0_{1}\nu_{e}$ (100) $\%$\\
$\tilde{e}_{R}$ & 209& $\tilde{\chi}^0_{1}e$ (100) $\%$\\
$\tilde{\tau}_{2}$ & 218& $\tilde{\chi}^0_{1}\tau$ (72) $\%$,\hspace{1 mm} $\tilde{\nu}_{\tau}W$ (27) $\%$\\
$\tilde{\chi}^0_{1}$ & 178& $\tilde{\nu}_{\tau}\nu_{\tau}$ (83) $\%$,\hspace{1 mm} $\tilde{\tau}^{\pm}\tau^{\mp}$ (17)$\%$\\
$\tilde{\tau}_{1}$ & 140&\\
$\tilde{\nu}_{\tau}$ & 73& $\tilde{G}\nu_{\tau}$\\
\hline
\end{tabular}
\end{center}
\caption{Spectrum and branching ratios for $M_{1/2}=500$ GeV, $\tan\beta=10$, $m^2_{H_{u}}=m^2_{H_{d}}=10^4\rm{GeV}^2$, $m_{0}=90$ GeV, $\Delta=3.57$
and $m_{h}=169$ GeV. Since the first and second generations are almost degenerate, we only give results for the first and third sfermion generations.}
\label{colltable1}
\end{table}

\begin{table}[htbp]
\begin{center}
\begin{tabular}{|c|c|c|}
\hline Sparticle & Mass[GeV] & Dominant decay modes\\ \hline $\tilde{g}$ & 1564 & $\tilde{q}_{L}q$ (16.2)$\%$,\hspace{1 mm}$\tilde{q}_{R}q$
(31.4)$\%$,\hspace{1 mm}$\tilde{b}_{1,2}b$ (20) $\%$,\hspace{1 mm} $\tilde{t}_{1}t$ (24) $\%$ \\ $\tilde{u}_{L},\tilde{d}_{L}$ & 1428, 1429 &
$\tilde{\chi}^0_{2}q$ (32) $\%$, \hspace{1 mm}$\tilde{\chi}^{\pm}_{1}q'$ (~64) $\%$\\ $\tilde{u}_{R},\tilde{d}_{R}$ & 1374,1368 & $\tilde{\chi}^0_{1}q$
(99) $\%$ \\ $\tilde{t}_{1}$ & 1112& $\tilde{\chi}^+_{1}b$ (19) $\%$,\hspace{1 mm}$\tilde{\chi}^0_{1}t$ (25) $\%$,\hspace{1 mm}$\tilde{\chi}^0_{3}t$
(17) $\%$,\hspace{1 mm}$\tilde{\chi}^+_{2}b$ (23) $\%$ \\ $H^{+}$ & 967& \\ $A$     & 946 &\\ $\tilde{\chi}^{0}_{4}$ & 864&
$\tilde{\chi}^{\pm}_{1}W^{\mp}$ (56) $\%$,\hspace{1 mm}$\tilde{\chi}^0_{2}h$ (19) $\%$\\ $\tilde{\chi}^{\pm}_{2}$ & 864& $\tilde{\chi}^0_{2}W^{\pm}$
(28) $\%$,\hspace{1 mm}$\tilde{\chi}^{\pm}_{1}Z$ (28) $\%$,\hspace{1 mm}$\tilde{\chi}^{\pm}_{1}h$ (20) $\%$\\ $\tilde{\chi}^0_{3}$ & 852&
$\tilde{\chi}^{\pm}_{1}W^{\mp}$ (56) $\%$,\hspace{1 mm}$\tilde{\chi}^0_{2}Z$ (26) $\%$\\ $\tilde{\chi}^0_{2}$ & 551& $\tilde{\nu}_{\tau}\nu_{\tau}$
(47) $\%$,\hspace{1 mm}$\tilde{\tau}^{\pm}_{1}\tau^{\mp}$ (39) $\%$\\ $\tilde{\chi}^{\pm}_{1}$ & 551& $\tilde{\nu}_{\tau}\tau^{\pm}$ (49)
$\%$,\hspace{1 mm}$\tilde{\tau}^{\pm}_{1}\nu_{\tau}$ (37) $\%$\\ $\tilde{e}_{L}$ & 486& $\tilde{\chi}^0_{1}e$ (100) $\%$\\ $\tilde{\nu}_{e}$ & 480&
$\tilde{\chi}^0_{1}\nu_{e}$ (100) $\%$\\ $\tilde{e}_{R}$ & 300& $\tilde{\chi}^0_{1}e$ (100) $\%$\\ $\tilde{\tau}_{2}$ & 303& $\tilde{\chi}^0_{1}\tau$
(72) $\%$,\hspace{1 mm}$\tilde{\nu}_{\tau}W$ (28) $\%$\\ $\tilde{\chi}^0_{1}$ & 249& $\tilde{\nu}_{\tau}\nu_{\tau}$ (90) $\%$,\hspace{1 mm}
$\tilde{\tau}_{1}^{\pm}\tau^{\mp}$(10) $\%$\\ $\tilde{\tau}_{1}$ & 217& $\tilde{\nu}_{\tau}W$ (100) $\%$\\ $\tilde{\nu}_{\tau}$ & 132&
$\tilde{G}\nu_{\tau}$\\ \hline
\end{tabular}
\end{center}
\caption{Spectrum and branching ratios for $M_{1/2}=700$ GeV, $\tan\beta=10$, $m^2_{H_{u}}=m^2_{H_{d}}=10^4\rm{GeV}^2$, $m_{0}=150$ GeV, $\Delta=6.13$
and $m_{h}=210$ GeV. Since the first and second generations are almost degenerate, we only give results for the first and third sfermion generations.}
\label{colltable2}
\end{table}
\begin{table}[htbp]
\begin{center}
\begin{tabular}{|c|c|c|}
\hline Sparticle & Mass[GeV] & Dominant decay modes\\ \hline $\tilde{g}$ & 829 & $\tilde{q}_{R}q$ (42)$\%$,\hspace{1 mm}$\tilde{b}_{1,2}b$ (16)
$\%$,\hspace{1 mm} $\tilde{t}_{1}t$ (42) $\%$ \\ $\tilde{u}_{L}$ & 853& $\tilde{\chi}^+_{2}q'$ (46) $\%$, \hspace{1 mm}$\tilde{\chi}^{0}_{4}q$ (23)
$\%$, \hspace{1 mm}$\tilde{\chi}^{+}_{1}q'$ (15) $\%$\\ $\tilde{d}_{L}$ &857 & $\tilde{\chi}^-_{2}q'$ (51) $\%$, \hspace{1 mm}$\tilde{\chi}^{0}_{4}q$
(25) $\%$\\ $\tilde{u}_{R},\tilde{d}_{R}$ & 750,737 & $\tilde{\chi}^0_{1}q$ (92) $\%$ \\ $H^{+}$ & 620& \\ $A$     & 588 &\\ $\tilde{t}_{1}$ & 539&
$\tilde{\chi}^+_{1}b$ (79) $\%$,\hspace{1 mm}$\tilde{\chi}^0_{1}t$ (20) $\%$\\ $\tilde{\chi}^{0}_{4}$ & 580& $\tilde{\chi}^{\pm}_{1}W^{\mp}$ (27)
$\%$,\hspace{1 mm}$\tilde{\tau}^{\pm}_{1}\tau^{\mp}$ (22) $\%$,\hspace{1 mm}$\tilde{\nu}_{\tau}\nu_{\tau}$ (30) $\%$\\ $\tilde{\chi}^{\pm}_{2}$ & 580&
$\tilde{\chi}^0_{2}W^{\pm}$ (13) $\%$,\hspace{1 mm}$\tilde{\chi}^{\pm}_{1}Z$ (28) $\%$,\hspace{1 mm}$\tilde{\chi}^0_{3}W^{\pm}$ (10) $\%$,\hspace{1
mm}$\tilde{\nu}_{\tau}\tau^{\pm}$ (28) $\%$,\hspace{1 mm}$\tilde{\tau}^{\pm}_{1}\nu_{\tau}$ (24) $\%$\\ $\tilde{\chi}^0_{3}$ & 400&
$\tilde{\chi}^0_{1}Z$ (88) $\%$\\ $\tilde{\chi}^0_{2}$ & 387& $\tilde{\tau}^{\pm}_{2}\tau^{\mp}$ (14) $\%$,\hspace{1
mm}$\tilde{\tau}^{\pm}_{1}\tau^{\mp}$ (54) $\%$\\ $\tilde{\chi}^{\pm}_{1}$ & 378& $\tilde{\nu}_{\tau}\tau^{\pm}$ (41) $\%$,\hspace{1
mm}$\tilde{\chi}^{0}_{1}W^{\pm}$ (47) $\%$\\ $\tilde{e}_{L}$ & 495& $\tilde{\chi}^0_{1}e$ (65) $\%$,\hspace{1 mm}$\tilde{\chi}^0_{2}e$ (20)
$\%$,\hspace{1 mm}$\tilde{\chi}^0_{2}e$ (20) $\%$,\hspace{1 mm}$\tilde{\chi}^{\pm}_{1}\nu_{e}$ (14) $\%$\\ $\tilde{\nu}_{e}$ & 489&
$\tilde{\chi}^0_{1}\nu_{e}$ (80) $\%$,\hspace{1 mm}$\tilde{\chi}^{\pm}_{1}e^{\mp}$ (18) $\%$\\ $\tilde{e}_{R}$ & 302& $\tilde{\chi}^0_{1}e$ (100)
$\%$\\ $\tilde{\tau}_{2}$ & 299& $\tilde{\chi}^0_{1}\tau$ (94) $\%$\\ $\tilde{\chi}^0_{1}$ & 249& $\tilde{\nu}_{\tau}\nu_{\tau}$ (99) $\%$\\
$\tilde{\tau}_{1}$ & 242& $\tilde{\nu}_{\tau}W$ (100) $\%$\\ $\tilde{\nu}_{\tau}$ & 163& $\tilde{G}\nu_{\tau}$\\ \hline
\end{tabular}
\end{center}
\caption{Spectrum and branching ratios for $M_{Y}=M_{2}=700$ GeV and $M_{3}=350$ GeV at the messenger scale, $\tan\beta=10$,
$m^2_{H_{u}}=m^2_{H_{d}}=1\times 10^4\rm{GeV}^2$, $m_{0}=150$ GeV, $\Delta=6.13$ and $m_{h}=208$ GeV. Since the first and second generations are almost degenerate, we only give results for the first and third sfermion generations.} \label{colltable3}
\end{table}
\begin{figure}[hbtp]
\centerline{\psfig{figure=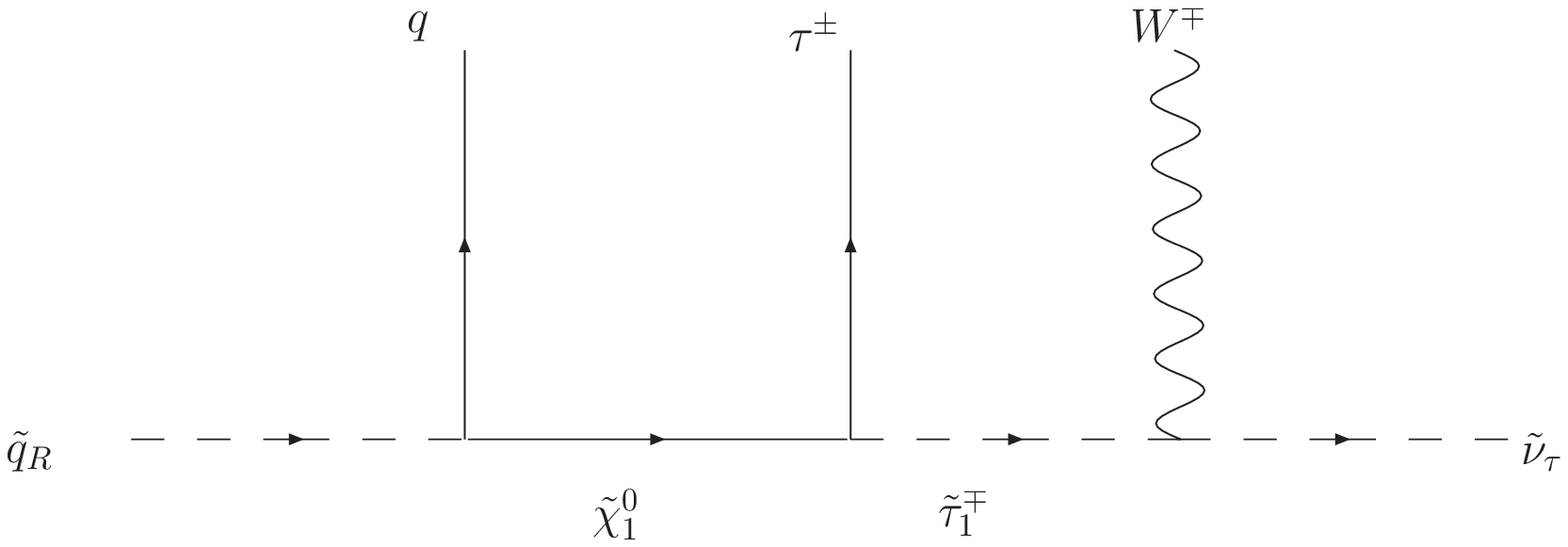,height=4.4 cm}}
 \vspace{0.2cm}
\centerline{\psfig{figure=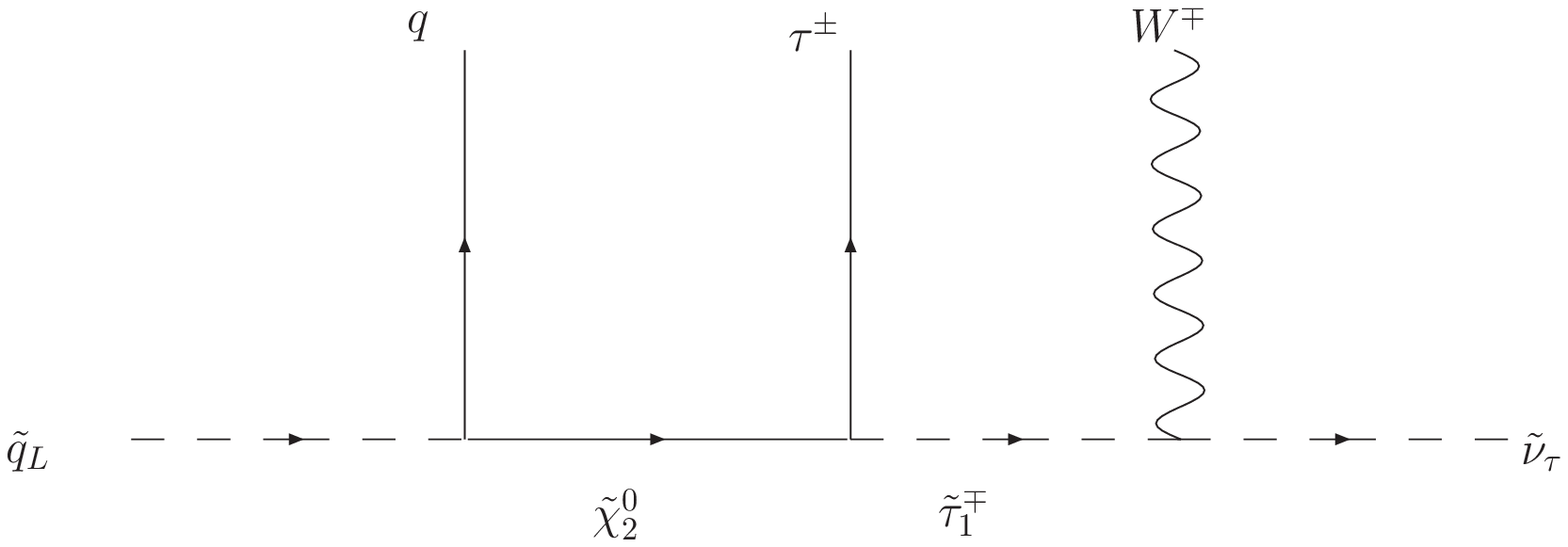,height=4.4 cm}}
 \vspace{0.2cm}
\centerline{\psfig{figure=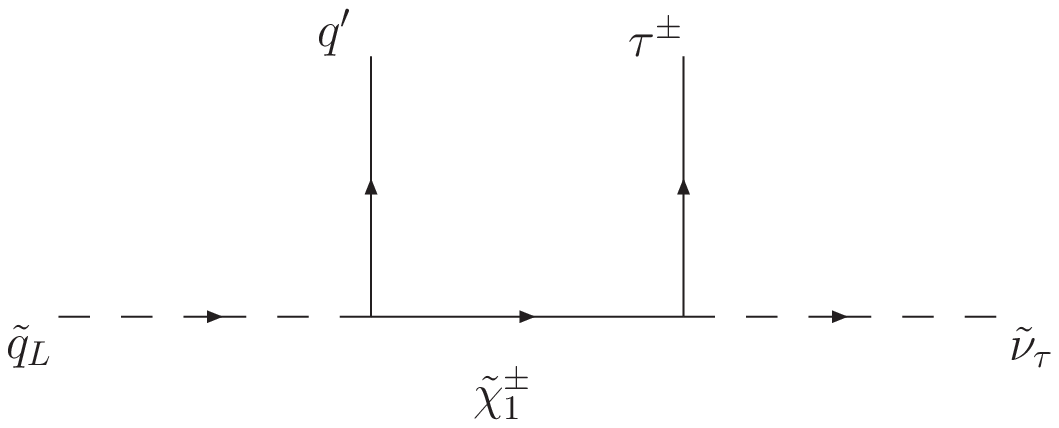,height=3.8 cm}}
\caption{ Examples of typical squark decays.}\label{chain1}
\end{figure}
Among the major differences with the case of non-universal Higgs
masses
(NUHM)~\cite{Ellis:2002iu}~\cite{Covi:2007xj},~\cite{Ellis:2008as}
is that the masses of the first two generation families run with
$SU(2)_{2}$ making left-handed sleptons and squarks heavier than
right handed ones, whereas in the NUHM case, for all generations,
since $S$ is positive, it tends to make left-handed sleptons of
all families relatively light. This is important since it affects
the decay of the charginos and the second lightest neutralinos,
which, contrary to the NUHM, now decay predominantly into only
third generation sleptons. The consequence is events with $\tau$
leptons in the final state, without a counterpart of similar
events with electrons or muons.

In the case of first and second generation squarks, seen in
Tables~\ref{colltable1} and \ref{colltable2}, as happens in the
MSSM, the right-handed squarks are still the lightest ones. This
leads, in the case of heavy gluinos, to smaller branching ratio of
gluinos into left-handed squarks than right-handed ones. This, in
turn, implies a bigger production of binos which tend to decay
more prominently into the sneutrino NLSP. Since the bino decay
products are invisible particles, the corresponding collider
signatures resemble those of the standard bino LSP scenarios. This
is partially compensated by the lighter third generation squarks,
which tend to decay into charginos and second lightest
neutralinos, which, in turn decay into third generation sleptons.
Moreover, as seen in Fig.~\ref{m0m1/2}, there are large regions of
parameter space where
$m_{\tilde{\chi}^1_{0}}>m_{\tilde{\tau}_{1}}$, giving a
non-negligible decay branching ratio of binos into lighter staus
which later decay into either on-shell or off-shell $W$'s. This is
particularly true for larger values of the Higgs mass. Examples of
typical decay cascades expected at the LHC or the Tevatron are
given in Fig.~\ref{chain1}. For these, we find, using
Table~\ref{colltable2}, that the total branching ratios are 10 \%,
12 \% and 31\%, respectively. In the case that the decays start
with a gluino, after considering the extra jet, the latter
branching ratios become $3.14\times 10^{-2}$, $1.94\times 10^{-2}$
and $5\times 10^{-2}$, respectively.

The existence of a heavy Higgs can lead to very interesting
events, with many $W^{\pm}$'s and bottom quarks in the final
state. This feature can be seen in Table~\ref{colltable2} where a
potential channel for Higgs production in supersymmetric particle
decays arises. This channel is given by the decay chain
$\tilde{g}\rightarrow \tilde{t}_{1}t\rightarrow
\tilde{\chi}^{+}_{2}bt\rightarrow \tilde{\chi}^{+}_{1}hbt$. In the
last stage of the decay chain we can have $\tilde{\chi}^{+}_{1}$
decaying either directly to the NLSP, $\tilde{\nu}_{\tau}\tau$,
about half the time or to $\tilde{\tau}_{1}\nu_{\tau}$ about 37 \%
of the time, the latter case being more phenomenologically
interesting since $\tilde{\tau}_{1}\rightarrow
W\tilde{\nu}_{\tau}$ produces on-shell $W^{\pm}$'s. Therefore, we
can have the following final states $\tilde{g}\rightarrow
hbWb\tilde{\nu}_{\tau}\tau$ and $\tilde{g}\rightarrow
hbWb\nu_{\tau}W\tilde{\nu}_{\tau}$ with gluino decay branching
ratios of $5.4\times 10^{-3}$ or $4.3\times 10^{-3}$,
respectively. Each gluino therefore may decay into a maximum of 2
bottom quarks and 4~$W^{\pm}$ states, plus missing energy. A pair
of $W^{\pm}$ should reconstruct the Higgs mass, while the
invariant mass of one of the bottom quarks and one of the
$W^{\pm}$'s should be equal to the top quark mass.

One can also consider the possibility of searching for the heavy
non-standard Higgs bosons, via their decay into SM-like Higgs
bosons. This was analyzed in Ref.~\cite{Cavicchia:2007dp} in a
similar supersymmetric model, in which the heavy Higgs boson
masses are induced by a large coupling to a singlet
field~\cite{Batra:2004vc},\cite{Barbieri:2006bg}. This search may
be possible, provided the non-standard Higgs decays into bottom
and tau pairs and/or the sneutrino decay branching ratio of
standard and non-standard Higgs bosons does not become too-large,
and is therefore more efficient for smaller values of $\tan\beta$
and/or larger sneutrino masses.

Table~\ref{colltable3} represents the case of a more compressed
spectrum, obtained by reducing the gluino mass with respect to the
Bino and Wino masses at the messenger scale~\cite{Martin:2007gf}.
This leads to the presence of relatively light stops, radically
changing the phenomenology, since now a large fraction of the
gluinos can decay into light stops and right-handed squarks and
negligibly into left-handed squarks. In addition, the lightest
stau becomes heavier than the second lightest neutralino and
lightest chargino, which are Higgsino-like and couple strongly to
the third generation squarks. This has the consequence of making
most decays end in either $\tilde{\chi}^0_{1}$ or
$\tilde{\nu}_{\tau}$ without passing through $\tilde{\tau}_{1}$.
Observe that when searching for stops at the LHC, the main decays
of $\tilde{t}_{1}$ in this sample point are 70\% into a bottom
quarks and a W gauge boson (including 20\% into an on-shell top
quarks) and missing energy and 30\% into a tau lepton, a bottom
quark and missing energy.

If we want to be in the region of parameter space where the tau
sneutrino is the NLSP, from Fig.~\ref{m0m1/2} and
Tables~\ref{colltable1}, \ref{colltable2} and \ref{colltable3}, we
see that the weak gauginos usually tend to be heavier than
sleptons in the low energy theory. Therefore in the case of
universal gaugino masses, we predict that the first and second
generation sleptons will mainly decay into the lightest
neutralino, that may be important for the associated direct
production phenomenology at either lepton or hadron colliders.
Observe that in Table~\ref{colltable3} due to the smallness of
$M_3$, the $\mu$ parameter is suppressed compared to the universal
gaugino case. Therefore, as mentioned above, the lightest chargino
as well as the second and third lightest neutralino are mainly
Higgsino-like.

We would like to conclude this subsection by considering the
effects of sfermion mixing in electroweak precision data. One of
the consequences of mixing is to lower the mass of the lightest
eigenstate with respect to the un-mixed state. Therefore, the
analysis presented in Fig.~\ref{m0m1/2} of the regions of
parameter space of consistent $\Delta T$, slightly underestimates
the region where the lightest $\tilde{\tau}$ is lighter than the
Bino. In the specific examples analyzed using SDECAY, given in
Tables~\ref{colltable1}, \ref{colltable2} and \ref{colltable3}, we
confirmed that the mixing does not change the general results
obtained in the case of zero mixing. We calculated the
contributions to $\Delta T$ coming from the mass splitting in
staus, third generation squarks and the charged Higgs sector.
These contributions are shown in Table~\ref{point1},
\begin{table}[htbp]
\begin{center}
\begin{tabular}{|c|c|c|c|c|c|c|}
\hline
$m_{h}$ [GeV] & $m_{0}$ [GeV]& $M_{1/2}$ [GeV] & $\Delta T_{\tilde{\tau}}$ & $\Delta T_{\tilde{Q}_{3}}$ & $\Delta T_{H^+}$ & $\Delta T_{tot}$\\
\hline
169 & 90 & 500 & $1.5\times 10^{-1}$ & $8.7\times 10^{-4}$ & $7.8\times 10^{-4}$ & $1.56\times 10^{-1}$ \\
\hline
210 & 150 & 700 & $1.9\times 10^{-1}$ & $2.6\times 10^{-3}$ & $8\times 10^{-3}$ & $2.03\times 10^{-1}$ \\
\hline
210 & 150 & (700,350) & $1.5\times 10^{-1}$ & $2.4\times 10^{-2}$ & $1.9\times 10^{-2}$ & $1.98\times 10^{-1}$ \\
\hline
\end{tabular}
\end{center}
\caption{Complete $\Delta T$ calculation for particular points in parameter space, which lie in the 68~\% confidence level ellipse in $\Delta
S$--$\Delta T$ plane.} \label{point1}
\end{table}
where we see that the results lie in the 68 \% confidence level
ellipses in the $\Delta S$--$\Delta T$ plane discussed in
Fig.~\ref{Fig.Ellipse}, and where in the third entry for the third
sample point we mean $M_{Y}=M_{2}=700$ GeV and $M_{3}=350$ GeV at
the messenger scale. Observe that in this last case, the third
generation squarks give a relatively large contribution to the $T$
parameter.

\section{Cosmological Constraints}

Within the supersymmetry breaking scheme described in the last
section, the tau-sneutrino is naturally the lightest SM
superpartner. However, as stressed before, if the messenger scale
is smaller than the standard GUT scale, the lightest
supersymmetric particle tends to be the gravitino. There could be
potentially constraints to this scenario specifically from the
observed Dark Matter (DM) density and from the decays of the
sneutrino or bino NLSP. First of all let us mention that in the
case of a long lived neutralino NLSP, obtained for a large
messenger scale of the order of $M_{GUT}$, the constraints coming
from its late decay into a photon and a gravitino are strong
enough to exclude all parameter space~\cite{Ellisneu}. However, if
the lightest neutralino is the LSP and not the gravitino, this
scenario could survive and give the correct DM relic density via
co-annihilation of the LSP with a light sneutrino as discussed
in~\cite{Kadota:2009vq}.

\begin{figure}[htb]
\centerline{\psfig{figure=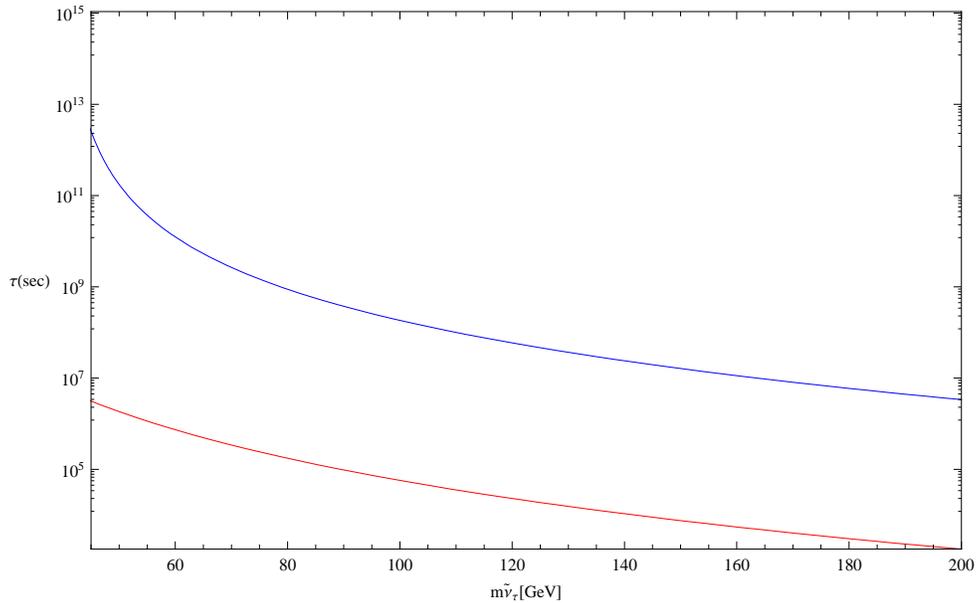,height=8 cm}} \caption{Lifetime of $\tilde{\nu}_{\tau}$ (sec) as a function of $m_{\tilde{\nu}_{\tau}}$
(GeV). The blue and red curves correspond to $m_{\tilde{G}}=40$ GeV and $m_{\tilde{G}}=1$ GeV, respectively}\label{lifetime}
\end{figure}

In the case of the sneutrino NLSP, the dominant decay channel is,
\begin{equation}
\tilde{\nu}_{\tau}\rightarrow \tilde{G}+\nu_{\tau}
\end{equation}
with a decay rate given by,
\begin{equation}
\Gamma_{\tilde{\nu}_{\tau}\rightarrow \tilde{G}+\nu_{\tau}}=\frac{m^5_{\tilde{\nu}_{\tau}}}{48\pi
M^2_{P}m^2_{\tilde{G}}}\left(1-\frac{m^2_{\tilde{G}}}{m^2_{\tilde{\nu}_{\tau}}}\right)^4.
\end{equation}
{}From this expression we can estimate the lifetime of the NLSP as
a function of the sneutrino and the gravitino masses. If the
gravitino mass is in the region $[1,100]$ GeV, as it is if the
messenger scale is of the order of the standard GUT scale, the
lifetime of the NLSP makes it potentially dangerous for BBN for
NLSP masses $\sim[45,400]$ GeV, the latter range being the
sneutrino masses of interest. Dependence of the sneutrino lifetime
on the gravitino mass can be seen in Fig.~\ref{lifetime}.
Assuming a gravitino LSP, its relic density is related to the
sneutrino NLSP density prior to decay by,
\begin{equation}
\Omega_{\tilde{G}}h^2=\frac{m_{\tilde{G}}}{m_{\tilde{\nu}_{\tau}}}\Omega_{\tilde{\nu}_{\tau}}h^2+\Omega^T_{\tilde{G}}h^2.\label{totdens}
\end{equation}
Given a sneutrino mass and assuming thermal equilibrium, its
freeze-out temperature is roughly $T\sim
m_{\tilde{\nu}_{\tau}}/25$. The relevant annihilation channels
have already been calculated in detail in the literature, (see for
instance, Ref.~\cite{Ellis:2002iu}), the most important being
$Z$-mediated s-channel annihilation to fermions. After a simple
calculation, one obtains, for a tau sneutrino mass of about a few
hundred~GeV, a thermal relic density,
$\Omega_{\tilde{\nu}_{\tau}}h^2\approx\mathcal{O}(10^{-3})$.

Since we are interested in the case
$m_{\tilde{G}}<m_{\tilde{\nu}_{\tau}}$, we see from
Eq.~(\ref{totdens}) and from the sneutrino relic density that in
order to obtain a gravitino DM relic density in accordance with
experiments, we must rely on either thermal or non-thermal
production of gravitinos at the reheating epoch. Gravitinos are
produced in the primordial plasma right after reheating from
scattering processes of the type, $x+y\rightarrow \psi+z$ where x,
y and z are relevant particles and $\psi$ is the gravitino. The
longitudinal component of the gravitino is related to the
goldstino, whose interaction with other particles is proportional
to $(m_{\tilde{G}}M_{P})^{-1}$. Hence, for masses
$m_{\tilde{G}}\gtrsim 100$ keV, gravitinos are not thermalized
with the rest of the plasma.

\begin{figure}[htb]
\centerline{ \psfig{figure=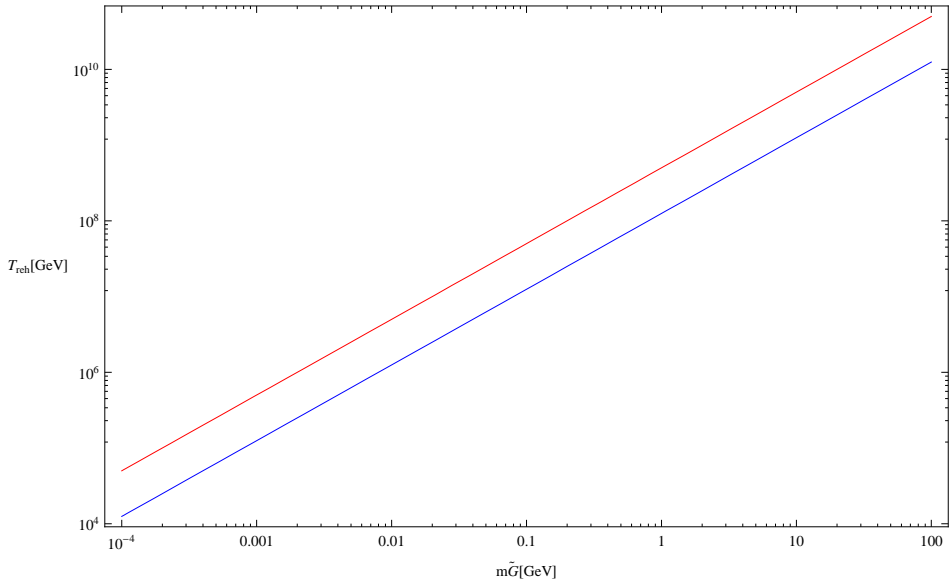,height=8cm,bb=0 0 450 286, clip=}}
\smallskip
\caption{ Reheating temperature, $T_{reh}$, as a function of the gravitino mass, $m_{\tilde{G}}$. The red and blue curves correspond to a soft gluino
mass $M_{3}=500,~1000$ GeV respectively.} \label{reh1}
\end{figure}

Assuming scattering processes to be the main source of gravitino
production, we can estimate, for a given gravitino mass, an upper
bound on the reheating temperature so that gravitinos are the main
component of DM. This was done in Ref.~\cite{de Gouvea:1997tn},
where solving the Boltzmann Equation using a thermally averaged
cross-section for scattering mainly off of gluinos and squarks, it
was found that,
\begin{equation}
\Omega_{\tilde{G}}\sim 0.2\times \left(\frac{m_{\tilde{G}}}{ \rm{GeV}}\right)^{-1}\times\left(\frac{M_{3}}{10^{3}\rm{GeV}}\right)^{2}
\times\left(\frac{T_{reh}}{10^8\rm{GeV}}\right).
\end{equation}
Taking into account that the gluinos are usually heavy, in
Fig.~\ref{reh1} we plotted the bound on the reheating temperature
as a function of the gravitino mass, assuming that the latter
saturates the DM component of the energy density in the Universe.

If the sneutrino decays during or after BBN, it could alter the
predictions for the light-element abundances. This is due mainly
to two effects. On one hand, if the mass difference between the
sneutrino and gravitino is sufficiently large, the sneutrino decay
produces high energy non-thermal neutrinos which, through
scattering processes with background particles, transfer part of
their energy to them. Some of the processes to consider are
$\nu_{\tau}+\bar{\nu}_{i,BG}\rightarrow
(e^{\pm},\mu^{\pm}\tau^{\pm})$,~
$\nu_{\tau}+\bar{\nu}_{i,BG}\rightarrow \pi^{+}+\pi^{-}$. On the
other hand, subdominant multi-body (bigger than two) sneutrino
decays into charged and/or strongly interacting particles can also
contribute significantly, despite having small branching rations.
Hadronic constraints, coming from processes such as,
\begin{equation}
\tilde{\nu}_{\tau}\rightarrow \tilde{G}\nu_{\tau}q\bar{q},\hspace{2mm}\tilde{\nu}_{\tau}\rightarrow \tilde{G}\nu_{\tau}Z,\hspace{2mm}
\tilde{\nu}_{\tau}\rightarrow \tilde{G}\tau W,
\end{equation}
with $W$ and $Z$ decaying into jets, turn out to be much stronger than electromagnetic constraints, coming from processes such as,
\begin{equation}
\tilde{\nu}_{\tau}\rightarrow \tilde{G}\nu_{\tau}l\bar{l},\hspace{2mm}\tilde{\nu}_{\tau}\rightarrow \tilde{G}\nu_{\tau}Z,\hspace{2mm}
\tilde{\nu}_{\tau}\rightarrow \tilde{G}\tau W,
\end{equation}
with $W$ and $Z$ decaying leptonically. The most stringent
constraint comes from the overproduction of D and $^{6}\rm{Li}$
produced in hadron showers induced by four-body
decays~\cite{Kanzaki:2006hm}.

There is also a potential constraint coming from the Cosmic
Microwave Background (CMB). For early decays, EM cascades are
thermalized completely by Compton scattering, double Compton
scattering and bremsstrahlung and no spectral distortion of the
CMB takes place. However for late decays, the spectrum cannot
relax to thermodynamic equilibrium, resulting in a Bose-Einstein
distribution with a non-vanishing chemical potential, $\mu_{pot}$.
The current constraint from experiments is $\mu_{pot} < 9\times
10^{-5}$. This constraint turns out to be negligible in the case
of a sneutrino NLSP with a gravitino LSP, as was shown
in~\cite{Feng:2004mt}.

All of these effects have been studied in
Refs.~\cite{Kanzaki:2006hm} and~\cite{Feng:2004mt} where, assuming
a thermal relic density for the sneutrino NLSP, it was found that
for a gravitino with a mass $m_{\tilde{G}}\gtrsim 10$ GeV, $\Delta
m= m_{\tilde{\nu}_{\tau}}-m_{\tilde{G}}\lesssim 290$ GeV and for
$m_{\tilde{G}}\lesssim 10$ GeV, there is no constraint on $\Delta
m$. As mentioned above, the biggest constraint comes from hadron
showers affecting D and $^{6}\rm{Li}$ BBN predictions. Therefore,
since in our model $m_{\tilde{\nu}_{\tau}} \lesssim 300$ GeV
(380~GeV) for $m_h < 270$~GeV ($m_h < 300$~GeV) (see
Fig.~\ref{sneutrino}) and we work with gravitino masses
1~GeV~$<m_{\tilde{G}}<100$~GeV, we conclude that it is only for
the larger Higgs masses that our parameter space gets slightly
truncated by cosmological constraints. Therefore most of our
parameter space of interest safely passes all cosmological
constraints.

\section{Model with Non-vanishing $SU(2)_1$ Gaugino Masses}

Throughout this article, we have assumed that the $SU(2)_1$
gaugino masses vanish at the messenger scale and therefore left
handed third generation soft SUSY breaking slepton and squark
masses as well as soft SUSY Higgs masses do not run under the
gauge group $SU(2)_{1}$. Hence, the corresponding particles are
not driven to large values via the $SU(2)_1$ gaugino corrections.
In particular third generation sleptons remain light as a
consequence of the transmission of SUSY breaking mechanism via
only $SU(3)_c\times SU(2)_2\times U(1)_Y$ gaugino masses.

We also saw how the doublet spectrum is split by $SU(2)_{1}$
D-terms in the effective low energy theory. This provides a
compensating contribution to the $T$-parameter that cancels the
logarithmic contribution coming from a large Higgs mass. Which
particle provides the cancellation to the standard Higgs
contribution to $\Delta T$ is a model dependent question that has
important consequences on the supersymmetric particle spectrum. In
most of this work, we have assumed third generation sleptons to be
light with respect to the third generation squarks and
non-standard Higgs doublet, and therefore to also provide the main
contribution to this cancellation. That the lightest doublet under
$SU(2)_{1}$ is the main contributor to $\Delta T$ is easy to
understand from looking at Eq.(\ref{deltaTsimple}), the squared
mass difference is independent of the soft masses and thus $\Delta
T$ is inversely proportional to the sparticle masses.

If we drop our assumption that the third generation sleptons,
squarks and Higgs soft masses do not run under $SU(2)_{1}$, then
their masses are pushed in an analogous fashion to larger values
at low energies by the strong $g_{1}$ coupling. For both third
generation squarks and sleptons there are small Yukawa effects
that lower their respective masses, but do not compensate the
strong $SU(2)_{1}$ and $SU(3)_c$ gaugino contributions. Therefore,
we conclude that in the case of $SU(3)_c\times SU(2)_{1}\times
SU(2)_{2}\times U_{Y}$ gaugino SUSY breaking mediation, it will be
mostly up to the non-standard Higgs doublet to provide the main
contribution to $\Delta T$. Although we will not make a detailed
phenomenological analysis, we will present the most important
phenomenological properties.

Let us analyze this case in slightly more depth. According to
Eq.~(\ref{mA}), the CP-odd Higgs mass depends only on the
difference of the soft supersymmetry breaking parameters of the
Higgs doublets $m_{H_{u,d}}^2$. Therefore, for universal soft
scalar masses, the CP-odd and charged Higgs become light in the
large $\tan\beta$ limit~\cite{Carena:1994bv}, when the top and
bottom Yukawa couplings affecting the Higgs mass parameters become
of the same order. Moreover, since now we have the contributions
from $SU(2)_1$ gauginos to the sparticle RGEs, right-handed
sfermions will tend to be lighter than left-handed ones. In
particular, due to the enhancement of the third generation slepton
and Higgs mass parameters, the right-handed stau will be affected
by large tau-Yukawa effects, pushing it to small values. Thus, the
phenomenology will be similar to the light stau NLSP in gaugino
mediation models. An example of the low energy spectrum arising
from this kind of models is presented in Table~\ref{Aspectrum}.
Note that large values of $\tan\beta$ and the $SU(2)_1$ gaugino
mass, $M_{1}$, may lead to dangerously small values of the
right-handed stau mass due to the negative Yukawa corrections
induced by the heavy left-handed sleptons. This is the main reason
why the value of $M_{1}$ in the example of Table~\ref{Aspectrum}
was chosen to be smaller than the rest of the gaugino masses.

Contrary to what happens in the case of vanishing $SU(2)_1$
gaugino masses, the light CP-odd scalars demanded in this scenario
are restricted by direct searches at the Tevatron collider and by
flavor constraints (see for example Ref.~\cite{Carena:2008ue}).
For instance, for values of $\tan\beta \simeq 50$, a light CP-odd
Higgs with a mass below 170~GeV is strongly constrained by current
Tevatron bounds~\cite{Tevatron}. This restricts an important
region of the parameter space corresponding to $m_h<180$ GeV as
can be observed from Fig.~\ref{sneutrino} by substituting the
sneutrino mass by the CP-odd Higgs mass. All Tevatron bounds may
be avoided for SM-like Higgs boson masses above 190~GeV, for which
heavier CP-odd Higgs bosons may still give a relevant $\Delta T$
contribution. Future searches at the Tevatron will probe the
existence of CP-odd Higgs bosons with masses up to 250~GeV for
this range of values of~$\tan\beta$, and will lead to further
tests of this scenario. Observe also that the flavor constraints
coming from the rare decays, $B_u \to \tau \nu$ and $b \to s
\gamma$, will be modified with respect to the MSSM case, due to
the larger mass difference between the charged and CP-odd Higgs
bosons. An analysis of this question, although very interesting,
is beyond the scope of the present article.

Cosmological constraints on this scenario arise in the case of a
stau NLSP with a gravitino LSP dark matter. These are related to
BBN constraints in the case of gravitino masses bigger than 100
KeV and to over-closure of the Universe and/or large scale
structure formation for smaller masses~\footnote{If very specific
conditions are fulfilled, however, stau NLSP scenarios may survive
and even serve to solve the so-called Lithium problem of Big Bang
Nucleosynthesis~\cite{Ellisneu,Bailly:2008yy}.}. For messenger
scales of the order of $M_{GUT}$, however, the gravitino can be
heavy enough so that the neutralino may be the lightest
supersymmetric particle, leading to the same cosmological and
phenomenological properties as the MSSM in the stau--neutralino
co-annihilation region. Although we will not explore this question
in more detail, the specific example of Table~\ref{Aspectrum}
shows a possibility in which the stau and neutralino are in a
range of masses such that efficient co-annihilation may take
place. Small variations of $m_0$ may lead to the precise value of
the stau-neutralino mass difference necessary to obtain the proper
relic density without affecting the required non-standard Higgs
spectrum.
\begin{table}[htbp]
\begin{center}
\begin{tabular}{|c|c|c|c|c|c|c|c|}
\hline $m_{h}$ [GeV] & $m_{A}$ [GeV]& $m_{H^+}$ [GeV] & $m_{\tilde{\tau}_{R}}$ [GeV] & $m_{\tilde{\tau}_{L}}$ [GeV]  & $m_{\tilde{\chi}^0_{1}}$ [GeV]&
$\mu$ [GeV] & $\Delta T_{tot}$\\ \hline 200 & 176 & 255 & $ 293 $ & $1020$ & $275$ & $321$ & $0.12$ \\ \hline
\end{tabular}
\end{center}
\caption{Partial spectrum for $\tan\beta=48$, $m_{0}=360$ GeV, $m^2_{H_{u}}=m^2_{H_{d}}=(200\rm{GeV})^2$, $M_{Y}=M_2=M_{3}=700$ GeV and $M_1=400$ GeV
at the messenger scale.} \label{Aspectrum}
\end{table}

\section{Conclusions}

In this article we have analyzed the phenomenological and
cosmological properties of the MSSM with enhanced $SU(2)$ D-terms.
Due to the splitting introduced in the third generation sfermion
and non-standard Higgs boson masses, we were able to raise the
Higgs mass up to about 300 GeV without being in conflict with
precision electroweak tests. We worked out a particular
supersymmetry breaking scenario that leads to the presence of
light third generation sleptons, which provide the necessary
positive contribution to the $T$ parameter to compensate for the
negative one induced by the heavy Higgs boson. Assuming universal
soft supersymmetry breaking gaugino, sfermion and Higgs masses at
the messenger scale, this supersymmetry breaking scenario leads to
the presence of heavy squarks and non-standard Higgs bosons. The
slepton spectrum is therefore determined from the requirement of
reestablishing agreement with precision electroweak data.

We studied the corresponding collider signatures, which, similar
to what happens in the light stau scenario in the MSSM, is
characterized by the presence of many tau's and copious missing
energy in the final states. For larger values of the Higgs mass,
the mass difference between the charged stau and the sneutrino is
large enough to allow two body decays of the staus into $W^{\pm}$
and $\tilde{\nu}$'s. Such final states would be a signature of the
enhanced $SU(2)$ D-term scenarios. Furthermore, Higgs searches
into pairs of $W$ and $Z$ gauge bosons would be similar to the SM
except in a small region of parameter space where tau sneutrinos
are light enough to allow Higgs decays into sneutrino pairs. This
additional decay mode may lead to a large enough modification of
the branching ratios of Higgs decay into gauge bosons to avoid
recent Tevatron bounds on the Higgs mass.

For the range of slepton masses consistent with precision
electroweak data, the sneutrinos cannot provide a consistent
thermal relic density. Indeed, mainly due to the $Z$-mediated
annihilation channel, the thermal sneutrino dark matter density
turns out to be too small. Therefore, we have assumed a gravitino
LSP, heavy enough to avoid thermal equilibrium with the plasma.
Gravitinos are produced by the decay of the light sneutrinos and
by the scattering of particles in the early universe. For
appropriate values of the reheating temperature the proper dark
matter relic density may be obtained.

In the presence of a gravitino LSP, the light sneutrino decays in
the early universe can lead to modifications of the light element
abundances. However, for sneutrino masses leading to agreement
with precision electroweak data and gravitino masses larger than a
few GeV, these constraints prove to be too weak to lead to any
relevant bound on the models discussed in this work.

Finally, we have discussed possible alternative scenarios in which
the $SU(2)_1$ gauginos are not suppressed and therefore the third
generation left-handed sleptons masses are driven to large values
by their re-normalization group effects. The non-standard Higgs
bosons will remain light for sufficiently large values of
$\tan\beta$ and may play a relevant role in canceling the negative
contributions to the $T$ parameter induced by the heavy Higgs.
This scenario is constrained by current Tevatron searches on
standard and non-standard Higgs bosons and by cosmological
constraints. Moreover, as is well known, the neutral and charged
Higgs components may induce relevant flavor violating effects in
such scenarios. However, the large mass splitting between the
neutral and charged components induced by the $SU(2)_1$ D-terms
leads to important differences in the flavor constraints with
respect to the traditional MSSM. In particular, the heaviness of
the charged Higgs dilutes the strong constraints on the MSSM
proceeding from the rare decays $B^+ \to \tau^+ \nu$ and $b \to s
\gamma$. It would be interesting to do a more complete analysis of
this question.

~\\ 
\section*{Acknowledgements}

We would like to thank P.~Batra, M.~Carena, A.~Delgado, E.~Ponton
and particularly T.~Tait for useful discussions and comments. This
work was supported in part by the DOE under Task TeV of contract
DE-FGO3-96-ER40956. Work at UC Davis was supported in part by U.S.
DOE grant No. DE-FG02-91ER40674. Work at ANL is supported in part
by the US DOE, Div.\ of HEP, Contract DE-AC02-06CH11357. Work of
N.~R.~Shah is supported by a Bloomenthal Research Fellowship.


\begin{thebibliography}{999}


\bibitem{Haber:1984rc} H.~P.~Nilles,
Phys.\ Rept.\  {\bf 110} (1984) 1;
H.~E.~Haber and G.~L.~Kane, 
Phys.\ Rept.\  {\bf 117}(1985) 75. 

\bibitem{Martin:1997ns}
S.~P.~Martin,
arXiv:hep-ph/9709356. 


\bibitem{Haber:1990aw}
  H.~E.~Haber and R.~Hempfling,
  Phys.\ Rev.\ Lett.\  {\bf 66}, 1815 (1991);
  Y.~Okada, M.~Yamaguchi and T.~Yanagida,
  Prog.\ Theor.\ Phys.\  {\bf 85}, 1 (1991);
  J.~R.~Ellis, G.~Ridolfi and F.~Zwirner,
  Phys.\ Lett.\  B {\bf 262}, 477 (1991).


\bibitem{Carena:1995bx}
  M.~S.~Carena, J.~R.~Espinosa, M.~Quiros and C.~E.~M.~Wagner,
  Phys.\ Lett.\  B {\bf 355}, 209 (1995)
  [arXiv:hep-ph/9504316];
  M.~S.~Carena, M.~Quiros and C.~E.~M.~Wagner,
  Nucl.\ Phys.\  B {\bf 461}, 407 (1996)
  [arXiv:hep-ph/9508343];
  H.~E.~Haber, R.~Hempfling and A.~H.~Hoang,
  Z.\ Phys.\  C {\bf 75}, 539 (1997)
  [arXiv:hep-ph/9609331];
  S.~Heinemeyer, W.~Hollik, and G.~Weiglein,
  Phys.\ Rev.\ D {\bf 58}, 091701 (1998) [arXiv:hep-ph/9803277];
  S.~Heinemeyer, W.~Hollik, and G.~Weiglein, 
  Phys.\ Lett.\ B {\bf 440}, 296 (1998) [arXiv:hep-ph/9807423]; 
  S.~Heinemeyer, W.~Hollik, and G.~Weiglein, 
  Eur.\ Phys.\ J.\ C {\bf 9}, 343 (1999)
  [arXiv:hep-ph/9812472]; 
  J.~R.~Espinosa and R.~J.~Zhang, 
  J. High Energy Phys. {\bf 0003}, 026 (2000) [arXiv:hep-ph/9912236]; 
  J.~R.~Espinosa and R.~J.~Zhang, 
  Nucl.\ Phys.\ B {\bf 586}, 3 (2000) [arXiv:hep-ph/0003246]; 
  M.~Carena, H.~E.~Haber, S.~Heinemeyer, W.~Hollik, C.~E.~M.~Wagner, and G.~Weiglein, 
  Nucl.\ Phys.\ B {\bf 580}, 29 (2000) [arXiv:hep-ph/0001002]; 
  G.~Degrassi, P.~Slavich, and F.~Zwirner, 
  Nucl.\ Phys.\ B {\bf 611}, 403(2001) [arXiv:hep-ph/0105096]; 
  A.~Brignole, G.~Degrassi, P.~Slavich, and F.~Zwirner, 
  arXiv:hep-ph/0112177; 
  S.~P.~Martin,
  Phys.\ Rev.\ D {\bf 67}, 095012 (2003) [arXiv:hep-ph/0211366]. 


\bibitem{Batra:2003nj}
  P.~Batra, A.~Delgado, D.~E.~Kaplan and T.~M.~P.~Tait, 
  JHEP {\bf 0402}, 043 (2004) [arXiv:hep-ph/0309149]; 


\bibitem{Batra:2004vc} P.~Batra, A.~Delgado, D.~E.~Kaplan and T.~M.~P.~Tait,
  JHEP {\bf 0406}, 032 (2004) [arXiv:hep-ph/0404251]. 

\bibitem{Bellazzini:2009ix}
   B.~Bellazzini, C.~Csaki, A.~Delgado and A.~Weiler,
   arXiv:0902.0015 [hep-ph].


\bibitem{Muller:1996dj} D.~J.~Muller and S.~Nandi, 
Phys.\ Lett.\ B {\bf 383}, 345 (1996) [arXiv:hep-ph/9602390];\\ 
E.~Malkawi, T.~Tait and C.~P.~Yuan, 
Phys.\ Lett.\ B {\bf 385}, 304 (1996) [arXiv:hep-ph/9603349];\\ 
E.~Malkawi and C.~P.~Yuan, 
Phys.\ Rev.\ D {\bf 61}, 015007 (2000) [arXiv:hep-ph/9906215]. 

\bibitem{Amsler:2008zzb}
  C.~Amsler {\it et al.}  [Particle Data Group],
  Phys.\ Lett.\  B {\bf 667}, 1 (2008).


\bibitem{Morrissey:2005uza}
  D.~E.~Morrissey, T.~M.~P.~Tait and C.~E.~M.~Wagner,
  Phys.\ Rev.\  D {\bf 72}, 095003 (2005)
  [arXiv:hep-ph/0508123].

\bibitem{Barbieri:2006bg}
  R.~Barbieri, L.~J.~Hall, Y.~Nomura and V.~S.~Rychkov,
  Phys.\ Rev.\  D {\bf 75}, 035007 (2007)
  [arXiv:hep-ph/0607332].


\bibitem{Barbieri:2006dq}
  R.~Barbieri, L.~J.~Hall and V.~S.~Rychkov,
  Phys.\ Rev.\  D {\bf 74}, 015007 (2006)
  [arXiv:hep-ph/0603188].


\bibitem{:2005ema}
    [The ALEPH, DELPHI, L3, OPAL, SLD Collaborations,
the LEP Electroweak Working Group, the SLD Electroweak and Heavy Flavour Groups ],
  Phys.\ Rept.\  {\bf 427}, 257 (2006)
  [arXiv:hep-ex/0509008].

\bibitem{HsearchesLHC}
  M.~Takakashi  [ATLAS Collaboration and CMS Collaboration],
  arXiv:0809.3224 [hep-ex].


\bibitem{HsearchesTeV}
  J.~Qian  [CDF and D0 Collaborations],
  arXiv:0812.3979 [hep-ex].
~\\ For the most recent Tevatron Higgs search results, see \\ http://theory.fnal.gov/jetp/talks/Verzocchi.pdf \\
http://theory.fnal.gov/jetp/talks/Jindariani.pdf \\

\bibitem{Maloney:2004rc}
  A.~Maloney, A.~Pierce and J.~G.~Wacker,
  JHEP {\bf 0606}, 034 (2006)
  [arXiv:hep-ph/0409127].

\bibitem{Ellis:2002iu}
  J.~R.~Ellis, T.~Falk, K.~A.~Olive and Y.~Santoso,
  Nucl.\ Phys.\  B {\bf 652}, 259 (2003)
  [arXiv:hep-ph/0210205].

\bibitem{Covi:2007xj}
  L.~Covi and S.~Kraml,
  JHEP {\bf 0708}, 015 (2007)
  [arXiv:hep-ph/0703130].

\bibitem{Ellis:2008as}
  J.~R.~Ellis, K.~A.~Olive and Y.~Santoso,
  JHEP {\bf 0810}, 005 (2008)
  [arXiv:0807.3736 [hep-ph]].

\bibitem{:chargino} LEPSUSYWG/01-03.1 and LEPSUSYWG/02-04.1, ALEPH, DELPHI, L3 and OPAL experiments
(http://lepsusy.web.cern.ch/lepsusy/Welcome.html).

\bibitem{Schmaltz:2000ei}
  M.~Schmaltz and W.~Skiba,
  Phys.\ Rev.\  D {\bf 62}, 095004 (2000)
  [arXiv:hep-ph/0004210].

\bibitem{Schmaltz:2000gy}
  M.~Schmaltz and W.~Skiba,
  Phys.\ Rev.\  D {\bf 62}, 095005 (2000)
  [arXiv:hep-ph/0001172].

\bibitem{Medina:2006hi}
  A.~D.~Medina and C.~E.~M.~Wagner,
  JHEP {\bf 0612}, 037 (2006)
  [arXiv:hep-ph/0609052].

\bibitem{Muhlleitner:2003vg}
  M.~Muhlleitner, A.~Djouadi and Y.~Mambrini,
  Comput.\ Phys.\ Commun.\  {\bf 168}, 46 (2005)
  [arXiv:hep-ph/0311167].


\bibitem{Cavicchia:2007dp}
  L.~Cavicchia, R.~Franceschini and V.~S.~Rychkov,
  Phys.\ Rev.\  D {\bf 77}, 055006 (2008)
  [arXiv:0710.5750 [hep-ph]].

\bibitem{Martin:2007gf}
  S.~P.~Martin,
  Phys.\ Rev.\  D {\bf 75}, 115005 (2007)
  [arXiv:hep-ph/0703097].

\bibitem{Ellisneu}
  R.~H.~Cyburt, J.~R.~Ellis, B.~D.~Fields, K.~A.~Olive and V.~C.~Spanos,
  JCAP {\bf 0611}, 014 (2006)
  [arXiv:astro-ph/0608562].

\bibitem{Bailly:2008yy}
  S.~Bailly, K.~Jedamzik and G.~Moultaka,
  arXiv:0812.0788 [hep-ph].

\bibitem{Kadota:2009vq}
  K.~Kadota, K.~A.~Olive and L.~Velasco-Sevilla,
  arXiv:0902.2510 [hep-ph].

\bibitem{de Gouvea:1997tn}
  A.~de Gouvea, T.~Moroi and H.~Murayama,
  Phys.\ Rev.\  D {\bf 56}, 1281 (1997)
  [arXiv:hep-ph/9701244].

\bibitem{Kanzaki:2006hm}
  T.~Kanzaki, M.~Kawasaki, K.~Kohri and T.~Moroi,
  Phys.\ Rev.\  D {\bf 75}, 025011 (2007)
  [arXiv:hep-ph/0609246].

\bibitem{Feng:2004mt}
  J.~L.~Feng, S.~Su and F.~Takayama,
  Phys.\ Rev.\  D {\bf 70}, 075019 (2004)
  [arXiv:hep-ph/0404231].





\bibitem{Carena:1994bv}
  M.~S.~Carena, M.~Olechowski, S.~Pokorski and C.~E.~M.~Wagner,
  Nucl.\ Phys.\  B {\bf 426}, 269 (1994)
  [arXiv:hep-ph/9402253].


\bibitem{Carena:2008ue}
  M.~Carena, A.~Menon and C.~E.~M.~Wagner,
  arXiv:0812.3594 [hep-ph].

\bibitem{Tevatron} http://www-cdf.fnal.gov/physics/new/hdg/results/htt\_070928/

http://www-d0.fnal.gov/Run2Physics/WWW/results/prelim/HIGGS/H29/H29.pdf


\end{thebibliography}
\end{document}